\documentclass{pasj00}

\begin{document}
\SetRunningHead{M. Mon et al.}{Spectroscopic Variations of EW Lac}
\Received{2013/02/20}
\Accepted{2013/03/05}

\title{Spectroscopic Variations of the Be-shell Star EW Lac in the V/R Variation Periods.}

%


\author{
Masahiro \textsc{Mon}\altaffilmark{1},
Masakazu \textsc{Suzuki}\altaffilmark{2},
Yuki \textsc{Moritani}\altaffilmark{3,4}
and
Tomokazu \textsc{Kogure},\altaffilmark{5}
}
\altaffiltext{1}{Osaka Shoin Wemen's University, 958 Sekiya, Kashiba, 639-0298}
\email{mon.masahiro@osaka-shoin.ac.jp}
\altaffiltext{2}{Kanazawa Institute of Technology, emeritus, 3-219 Onuka, Kanazawa, 921-8147}
\altaffiltext{3}{Department of Astronomy, Faculty of Science, Kyoto University, Sakyo-ku, Kyoto, 606-8502}
\altaffiltext{4}{Hiroshima Astrophysical Science Center, Hiroshima University, 1-3-1 Kagamiyama, Higashi-Hiroshima, 739-8526}
\altaffiltext{5}{Kyoto University, emeritus, 1-10 Toganoo, Hashimoto, Yawata, Kyoto, 614-8322}


\maketitle

\begin{abstract}
EW Lac has shown remarkable V/R variations in 1976 -- 1986, and a similar V/R variation has started again in around 2007, after some quasi-periodic V/R variation.
For the first V/R variation event, we analyze the spectroscopic behaviors of emission lines and shell absorption lines of the Balmer series.
The V/R variations of the H$\alpha$ through H$\delta$ lines are characterized by the different manner of variations of the individual emission lines in time lag and in duration of the V/R phases.
Weak correlation is also notable between the V/R variations and other variations of the line-profile parameter such as the peak velocities, emission line intensities and peak separations.
We analyze shell absorption lines for higher members of the Balmer series for their central depths and radial velocities.
The optical depth of the H$\alpha$ line is in a range of 2000 to 6000, and its long-term variation discloses different behaviors as compared to the V/R variations.
Combining the analyses of emission and shell absorption lines, and regarding the V/R variation as wave propagation phenomena, we find for the 1976 -- 1986 event that the V/R variation is of retrograde structure and a spiral structure is likely formed inside the disk in the latter half of this event.
The weak correlations among physical parameters are suggestive of the disk truncated at some radius.
It is noticed that remarkable stellar brightening has occurred in the latter half of the event, accompanying marked decrease of emission line intensities.
As for the V/R variation appeared in around 2007, which looks like a recurrence of the previous event, we found less developed state of the disk without appreciable time lag.
\end{abstract}

\section{Introduction}
Be stars are the stars that show emission lines in the Balmer lines, singly ionized metal lines and sometimes neutral helium lines in their optical spectra.
It is now widely accepted that Be stars are rapidly rotating stars surrounded by disks or rings in the low-latitude regions, where the emission lines are formed.
Be stars are classified into ordinary Be stars and Be-shell stars according to whether sharp absorption lines, so-called shell absorption lines, exist or not in the Balmer and metallic lines.
This classification is somehow related with the inclination of the Be disk.
Namely, Be-shell stars are considered to have a disk nearly edge-on from the observer.
Be stars usually exhibit irregular variations in both photometric and spectral features on various timescales from less than one day to several decades.
Phase change among B, Be, and Be-shell stars are occasionally observed.
A general review on Be stars can be seen in \citet{PR03}.
\citet{MZHGB98} reviewed long-term variations in various Be stars.
On the formation of Balmer emission lines and shell absorption lines, an overview is seen in \citet{KL07}.

EW Lac (HD217050, B4IIIep) has been known as a typical Be-shell star since 1930's, showing irregular variations in the shell lines \citep{K75,P80,HFCC87}.
Remarkable V/R variations of the Balmer emission lines have been observed in the period from 1976 through 1986, and a similar variation may have started in around 2007.

For the 1976 -- 1986 events, several observations have been carried out.
\citet{KS84} examined the V/R ratios for the H$\alpha$ through the H$\delta$ lines in the period 1960 -- 1983, based on visual inspection of spectrograms, and  qualitatively showed that the V/R variations in 1975 -- 1983 proceeded  accompanying some time lag for different Balmer lines.
\citet{SK86} showed the V/R variations for the H$\alpha$ and the H$\beta$ lines in the period from 1970 up to 1985, but without mentioning the time lag.
\citet{HFCC87} divided the spectroscopic history of EW Lac into quiet (1960 -- 1974) and active (1975 -- 1984) phases, and compared the spectroscopic features of these phases.
The quiet phase corresponds to the Be-shell star state and active phase to the remarkable V/R variation state.
They examined the Balmer lines in the active phase and showed the existence of phase lags in the V/R variations for different Balmer lines and also in the relation between radial velocities of the emission-line centers and a higher Balmer line (H20).

\citet{K75} has analyzed the shell absorption lines and shown that the circumstellar disk of this star has large optical depth of the H$\alpha$ line, as large as 2000 to 6000, accompanying some irregular variation in the period 1961 -- 1972.
Shell absorption lines in the Balmer series with higher quantum number (higher Balmer members) are generally optically thin and can provide information on the internal structure of the circumstellar disk in front of the photosphere, such as optical depths, effective disk area, gas motions and electron density.

In this paper we consider the spectroscopic behaviors of the Balmer emission lines and shell absorption lines, mostly in the first V/R variation period 1976 -- 1986.
We consider only long-term variations much longer than one day.
Although it is interesting to see that EW Lac has shown the short-term V/R variations less than one day \citep{PS90,FHHMZGHJKLSTN00}, such a short-term variability is beyond this work. 

In section \ref{sec:ObsData}, we present the observational data and reduction procedure.
Section \ref{sec:SpecVar} is devoted to present the spectroscopic variations of emission line profiles in the Balmer series, including V/R ratios, peak separations and central core components.
In section \ref{sec:VoverR2000s} we show the new appearance of the V/R variation in around 2007.
In sections \ref{sec:ShellVar} and \ref{sec:RVofShell}, we analyze Balmer shell absorption lines for the line intensities and radial velocities, and investigate the internal structure of the disk.
We derive some physical parameters in section \ref{sec:Paras}.
In section \ref{sec:Photo}, we examine the relationship between photometric variations and V/R variations.
Structure of the disk related to the V/R variations is discussed in section \ref{sec:Diss}.
We summarize this paper in section \ref{sec:Summ}.

\section{Observational Data and Measurements of Line-Profile Parameters}\label{sec:ObsData}
\subsection{Observational Data and Reduction Procedure}\label{subsec:ObsRed}
In this section we present the observational data obtained in 1966 -- 1986.
We have measured 195 photographic spectrograms of EW Lac, obtained by the Coud\'{e} spectrograph attached to the 188 cm reflector at the Okayama Astrophysical Observatory (OAO).
The plate list is shown partly in table \ref{tab:PlList} (Full table is given in the electronic version). 
The list contains the observation journal, plate number and the names of Balmer lines contained in each plate.
The spectral dispersion is 10 \AA/mm (violet) and 20 \AA/mm (H$\alpha$) for C4-plates and 4 \AA/mm (violet) and 8 \AA/mm (H$\alpha$) for C10-plates, respectively.

All of the spectrograms were digitized with the PDS microdensitometer at the Kwasan Observatory of Kyoto University.
The data analysis was carried out by the program developed by M. Suzuki, by making use of the computers in Kyoto University and Kanazawa Institute of Technology.
The results were recorded on magnetic tape in the form of relative intensity traces in wavelength scale corrected for the solar motion.
We have measured the parameters characterizing the line profiles on these tracings in the way described in the following subsections.

\begin{table*}
  \caption{The plate list  of EW Lac. Full table is given in the electronic version.}\label{tab:PlList}
  \begin{center}
    \begin{tabular}{ccccccccc}
      \hline
Date & JD2400000$+$ & Plate No. & H$\alpha$ & H$\beta$ & H$\gamma$ & H$\delta$ & H$\epsilon$ - H35 \\
\hline \hline
1966/08/22 & 39359.143 & C4-1461 & & $\bigcirc$ & & & \\    
1966/08/22 & 39359.210 & C4-1464 & & & & $\bigcirc$ & $\bigcirc$ \\
1966/08/22 & 39359.220 & C4-1465 & & & $\bigcirc$ & $\bigcirc$ & \\   
1969/08/29 & 40462.164 & C4-2362 & & & $\bigcirc$ & $\bigcirc$ & \\    
1969/08/29 & 40462.233 & C4-2363 & & & & $\bigcirc$ & $\bigcirc$ \\
1969/12/22 & 40577.893 & C4-2437 & & & & $\bigcirc$ & $\bigcirc$ \\
1969/12/22 & 40577.931 & C4-2438 & & & $\bigcirc$ & $\bigcirc$ & \\    
1969/12/22 & 40577.962 & C4-2439 & & $\bigcirc$ & & & \\    
1969/12/22 & 40577.998 & C4-2440 & $\bigcirc$ & & & & \\    
      \hline
    \end{tabular}
  \end{center}
\end{table*}

\subsection{Subtraction of Emission Lines}\label{subsec:SubEm}
We have selected 10 spectrograms with better {$S/N$} ratio among OAO plates, and prepared the average line profiles for the Balmer lines.
These profiles were used for the fitting with Krucz's model profiles broadened by stellar rotation, under the parameter ranges of $T_{\mathrm{eff}}$ (16000 -- 18000 K), $\log$ $g$ (2.5 -- 3.5) and $V{\mathrm{sin}}i$ (250 -- 450 $\mathrm{km\;s^{-1}}$) (\cite{K79}).
We have thus adopted the following parameter set as the best fitting with the OAO profiles in the wing parts.

$T_{\mathrm{eff}}$ = 16000 K, $\log$ $g$ = 3.0 and $V{\mathrm{sin}}i$ = 280 $\mathrm{km\;s^{-1}}$     

\begin{figure*}
  \begin{center}
    \FigureFile(150mm,110mm){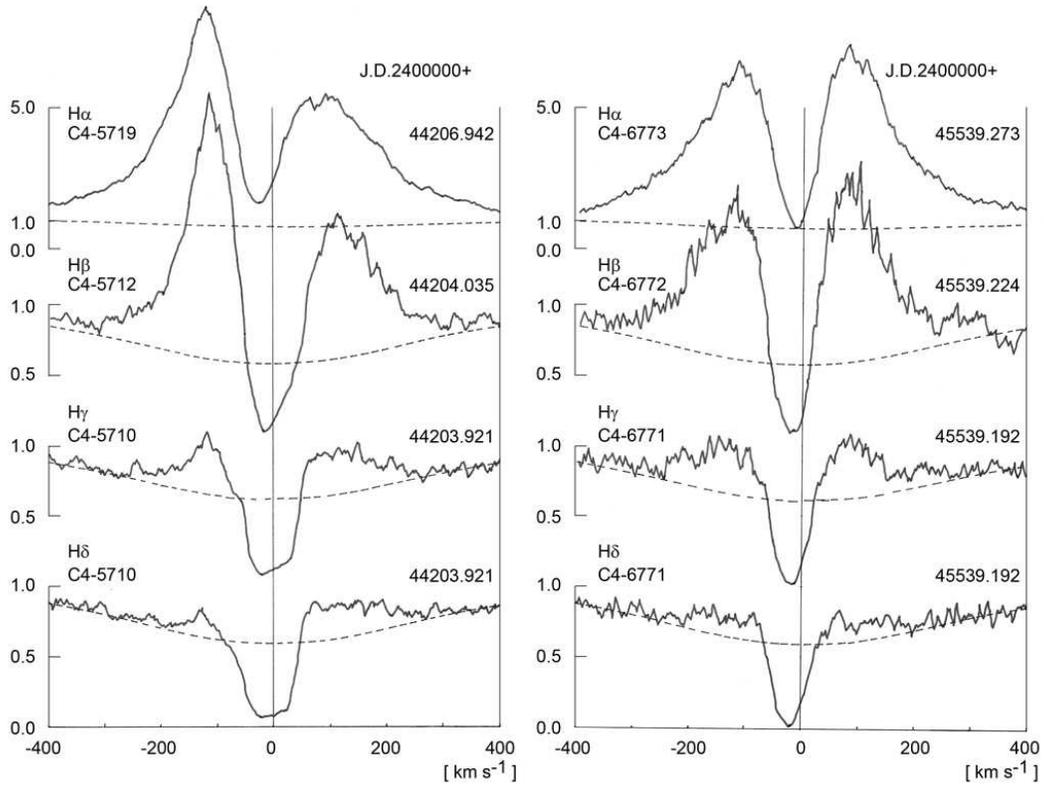}
  \end{center}
  \caption{
Line profiles of the H$\alpha$ through the H$\delta$.
Left panel shows the profiles in 1979 November and right panel those in 1983 June.
Plate number and Julian Day are written above each of the profiles.
Dashed lines indicate the photospheric lines taken from broadened Kurucz's model.
The vertical scale is 5 times smaller for the H$\alpha$ line than for the other lines.
}\label{fig:SamProf}
\end{figure*}

In addition, we have measured the half-width at half-maximum $\Delta_{1/2}$ of He I $\lambda$4471 for 10 selected OAO plates. $V{\mathrm{sin}}i$ is estimated by making use of the correlation between $V{\mathrm{sin}}i$ and $\Delta_{1/2}$ for several stars which show the parameters nearer to our parameter set. The data are taken from \citet{CZBMCG01}. The result is that $V{\mathrm{sin}}i$ = 279 $\pm$ 15 $\mathrm{km\;s^{-1}}$, which is in accord with the above parameter set. The derived parameter set is also in accord with the spectral type B4IIIpe, which is given by \citet{PHBKHKRS97}, for instance. Notice that our parameter set is used for determining the underlying photospheric profiles and is not sensitive to the discussions of the present paper. Samples of the profiles of the Balmer lines H$\alpha$ through H$\delta$, and the photospheric absorption lines based on this parameter set are shown in figure \ref{fig:SamProf}.

\subsection{Measurements of Line-Profile Parameters}\label{subsec:MeasureProf}
\begin{figure*}
  \begin{center}
    \FigureFile(150mm,60mm){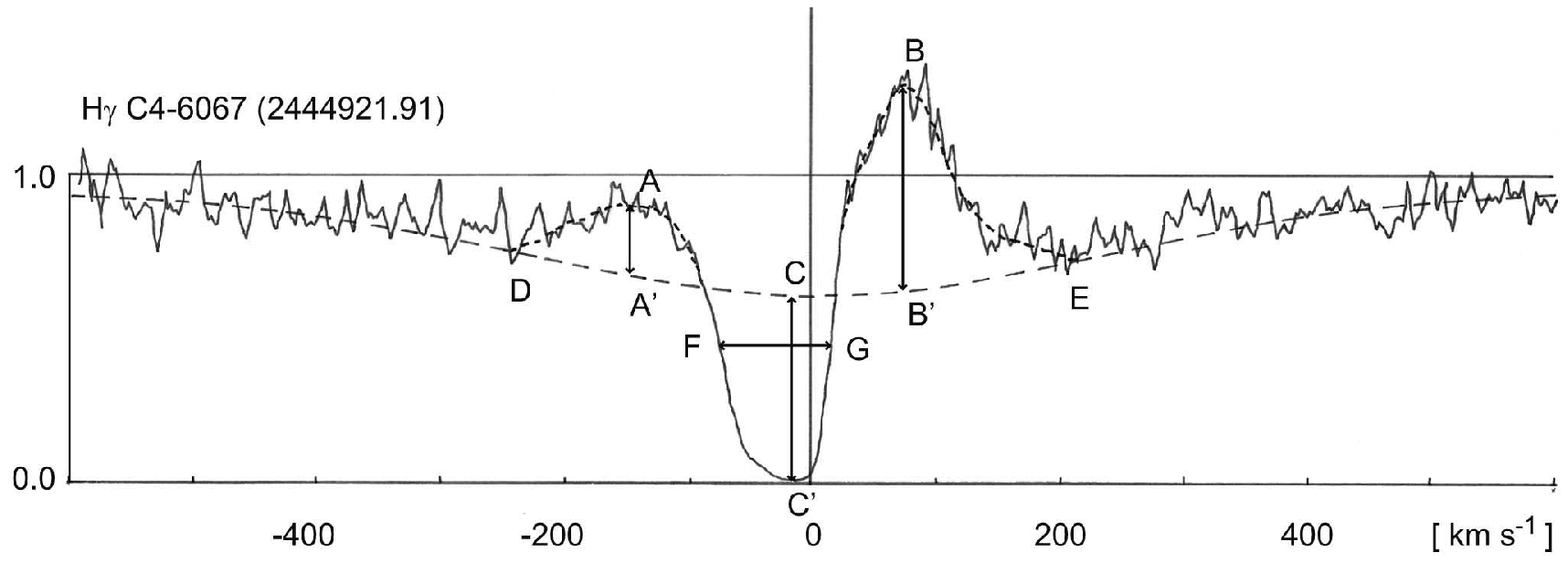}
  \end{center}
  \caption{Definition for the profile measurements.}\label{fig:DefMeasure}
\end{figure*}

In all of the observed Balmer emission line profiles, we have measured the following line-profile parameters which are defined in figure \ref{fig:DefMeasure}.

\begin{enumerate}
\item Heights of emission peaks $I_{v}$ (AA') and $I_{r}$ (BB') for the violet and red components, respectively, in unit of nearby continuum, then the V/R ratio is defined by the ratio $I_{v}$/$I_{r}$.
\item Heliocentric radial velocities of the emission peaks, $V_{v}$ at A, and $V_{r}$ at B in $\mathrm{km\;s^{-1}}$.
\item Peak separation 2$\Delta_p$ (velocity difference between A and B) of emission lines in $\mathrm{km\;s^{-1}}$.
\item Equivalent width of the emission component above the photospheric line $EW_e$ in \AA.
\item Heliocentric radial velocities of the deepest point $V_{c}$ (at C) in $\mathrm{km\;s^{-1}}$.
\item Depth of central core absorption measured from the photospheric absorption $D_c$ (CC').
\item Full width at half-intensity of central core absorption 2$\Delta_s$ (velocity difference between F and G) in $\mathrm{km\;s^{-1}}$.
\end{enumerate}

The numerical data of these parameters are partly shown in table \ref{tab:ValMeasure}, where the mean values of  each observational time are given (Full table is given in the electronic version).

\begin{table*}
  \caption{The measured quantities of emission line profiles. Full table is given in the electronic version.}\label{tab:ValMeasure}
  \begin{center}

    \begin{tabular}{rcccrrrrrrr}
      \hline
JD2400000$+$ & $I_v$ & $I_r$ & V/R & $V_v$ & $V_r$ & 2$\Delta_p$ & $EW_e$ & $V_c$ & $D_c$ & 2$\Delta_s$ \\
\hline \hline
H$\alpha$ & & & & & & & & & & \\
40578 & 5.83 & 5.70 & 1.02  & $-$80.0 & 107.5 & 187.5 & 58.17 & 15.0 & 1.38  & \\
40868 & 5.20 & 5.25 & 0.99  & $-$112.5 & 110.0 & 222.5 & 50.87 & 5.0 & 1.25  & \\
41195 & 3.45 & 3.40 & 1.01  & $-$130.0 & 95.0 & 225.0 & 35.56 & $-$16.0 & 0.25  & \\
H$\beta$ & & & & & & & & & & \\
39359 & 0.94 & 1.02 & 0.92  & $-$135.0 & 92.5 & 227.5 & 3.61 & $-$21.5 & 0.54  & 64.5 \\
40580 & 0.82 & 0.77 & 1.07  & $-$130.0 & 90.0 & 220.0 & 4.71 & $-$20.0 & 0.25  & 41.3 \\
41195 & 0.60 & 0.64 & 0.94  & $-$135.0 & 100.0 & 235.0 & 3.25 & $-$24.0 & 0.25  & 59.0 \\
H$\gamma$ & & & & & & & & & & \\
39359 & 0.24 & 0.32 & 0.75  & $-$125.0 & 110.0 & 235.0 & 1.20 & $-$13.0 & 0.60  & 65.5 \\
40462 & 0.28 & 0.36 & 0.78  & $-$125.0 & 105.0 & 230.0 & 1.10 & $-$11.5 & 0.56  & 58.5 \\
40578 & 0.26 & 0.28 & 0.93  & $-$125.0 & 101.3 & 226.3 & 1.06 & $-$16.0 & 0.52  & 49.3 \\
H$\delta$ & & & & & & & & & & \\
39359 & 0.13 & 0.17 & 0.77  & $-$122.5 & 95.0 & 217.5 & 0.70 & $-$22.0 & 0.58  & 60.5 \\
40462 & 0.11 & 0.13 & 0.85  & $-$125.0 & 97.5 & 222.5 & 0.62 & $-$15.5 & 0.53  & 57.8 \\
40578 & 0.15 & 0.16 & 0.94  & $-$125.0 & 101.3 & 226.3 & 0.57 & $-$12.0 & 0.53  & 60.4 \\
      \hline
    \end{tabular}
  \end{center}
\end{table*}

\section{Spectroscopic Variations of the Balmer Emission Lines}\label{sec:SpecVar}
\begin{figure*}
  \begin{center}
    \FigureFile(150mm,130mm){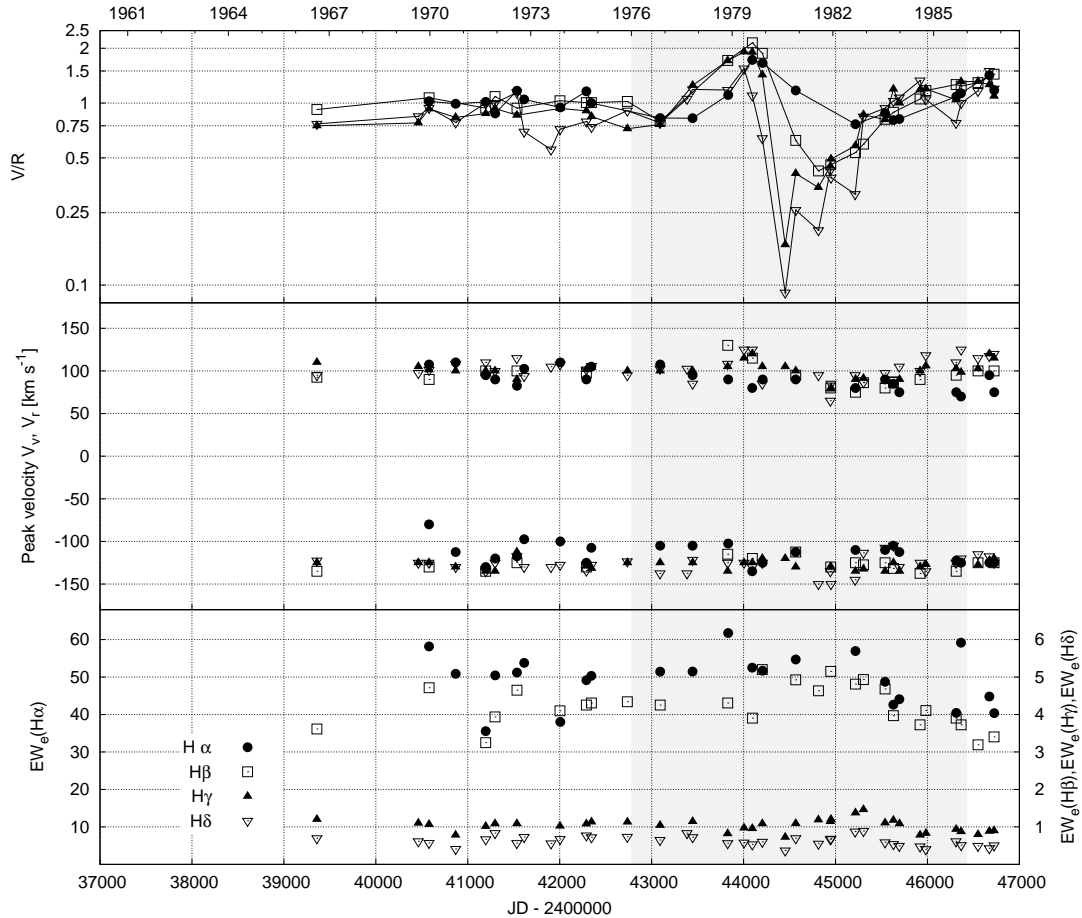}
  \end{center}
  \caption{
{\it Top panel} : V/R variations for the H$\alpha$ through the H$\delta$ lines.
{\it Middle panel} : Radial velocities of the violet and the red peaks of emission lines.
{\it  Bottom panel} : Equivalent widths of emission component above the underlying photospheric absorption.
The gray region indicates a period of the V/R variation (1976 -- 1986).
}\label{fig:VoRPeakEW}
\end{figure*}

In this section we present the observed features of variation in the line profiles of EW Lac.
In figure \ref{fig:VoRPeakEW} we show the long-term variations of V/R ratios, heliocentric peak velocities ($V_{v}$ and $V_{r}$) and emission equivalent widths ($EW_e$) for the H$\alpha$ through the H$\delta$ lines.
Main features are described in the following subsections.

\subsection{V/R Variations}\label{subsec:VoRVar}
Although the V/R variations seem to have started in early 1970's in the H$\gamma$ and the H$\delta$ lines with shallow V $<$ R state (V/R $\sim$ 0.8 -- 0.9), the V/R variations become notable in 1976 and continued to around 1986 as seen in figure \ref{fig:VoRPeakEW} (gray region).
The V/R variation after 1986 will be considered in subsection \ref{subsec:Long_VoverR}.
The remarkable V/R variation may be characterized by the following features:

\begin{enumerate}
\item The epoch of V/R maximum or V/R minimum depend on the lines.
That is, the V/R maximum appeared almost at the same epoch within 100 days in the middle of 1979 for all emission lines, whereas, each line reached the V/R minimum at different time.
The H$\delta$ and H$\gamma$ lines reached it earlier, and then the H$\beta$ and H$\alpha$ lines followed in this order. 
The epoch of V/R minimum in the H$\alpha$ was delayed for approximately 600 days, as compared with that in the H$\delta$ line (see figure \ref{fig:VoRPeakEW} and table \ref{tab:DurVoR}).
Hereafter, we call this delay the time lag of the V/R variation.
The time lag became evident after the epoch of V/R maximum phase in 1979.
\item The durations of V/R phase (V $>$ R or V $<$ R phase) differ considerably among the lines.
The approximate durations are shown in table \ref{tab:DurVoR}.
One may see that the total duration gradually declined from the H$\alpha$ toward higher members.
\item The range of V/R variation are moderate for the H$\alpha$ and the H$\beta$ lines from V/R $=$ 0.4 to 2.1, whereas those for the H$\gamma$ and the H$\delta$ lines are large from V/R $=$ 0.09 to 2.0. 
For the latter lines, small violet emission component is remarkable at V/R minimum phase.
\end{enumerate}

\begin{table*}
  \caption{Durations of the phase of V/R variation in days.}\label{tab:DurVoR}
  \begin{center}
    \begin{tabular}{ccccccccccc}
      \hline
Phase & V/R = 1 & V $>$ R & ${\mathrm{V/R}}_{max}$ & V $>$ R & V/R = 1 & V $<$ R & ${\mathrm{V/R}}_{min}$ & V $<$ R & V/R = 1 & Total \\
\hline \hline
H$\alpha$ & & 480 & & 710 & & 680 & & 950 & &2820 \\
H$\beta$ & & 820 & &  370 & & 710 & & 610 & & 2510 \\
H$\gamma$ & & 810 & & 140 & & 580 & & 780 & & 2310 \\
H$\delta$ & & 710 & & 120 & & 310 & & 950 & & 2090 \\
      \hline
    \end{tabular}
  \end{center}
\end{table*}

\subsection{Peak Velocities and Emission Line Intensities}\label{subsec:PvEl}
In figure \ref{fig:VoRPeakEW} are also shown the variations of heliocentric radial velocities at the violet and the red peaks (middle panel) and emission equivalent widths (bottom panel).

Weak correlation of the peak velocities $V_v$ and $V_r$ with the V/R variations is apparent.
The radial velocities of the red peak slightly shifted red-ward along with the increase of the V/R value, whereas the violet components indicate almost no relationship.
As a whole, the correlation is week.
The average velocities and their standard deviations are given in table \ref{tab:VelEP}.
The standard deviation is less than around 10 $\mathrm{km\;s^{-1}}$.

The equivalent widths of emission lines seem to show weak correlation with the V/R variations, though some large scatter is seen.
The average equivalent widths in the V/R variation period are as follows:

$EW_e$ [\AA ] = 50.1 (H$\alpha$), 4.3 (H$\beta$), 1.0 (H$\gamma$) and 0.67 (H$\delta$). 

\begin{table*}
  \caption{Average radial velocities of the emission peaks and standard deviations [$\mathrm{km\;s^{-1}}$].}\label{tab:VelEP}
  \begin{center}
    \begin{tabular}{ccccc}
      \hline
Line & H$\alpha$ & H$\beta$ & H$\gamma$ & H$\delta$ \\
\hline \hline
$<V_v>$ & $-$113.5 $\pm$ 12.3 & $-$127.3 $\pm$ 6.7 & $-$127.4 $\pm$ 5.3 & $-$127.2 $\pm$ 9.9 \\
$<V_r>$ & 90.9 $\pm$ 11.9 & 94.9 $\pm$ 12.8 & 100.2 $\pm$ 10.0 & 102.7 $\pm$ 13.0 \\
      \hline
    \end{tabular}
  \end{center}
\end{table*}

\subsection{Peak Separations}\label{subsec:PeakSep}
\begin{figure*}
  \begin{center}
    \FigureFile(150mm,70mm){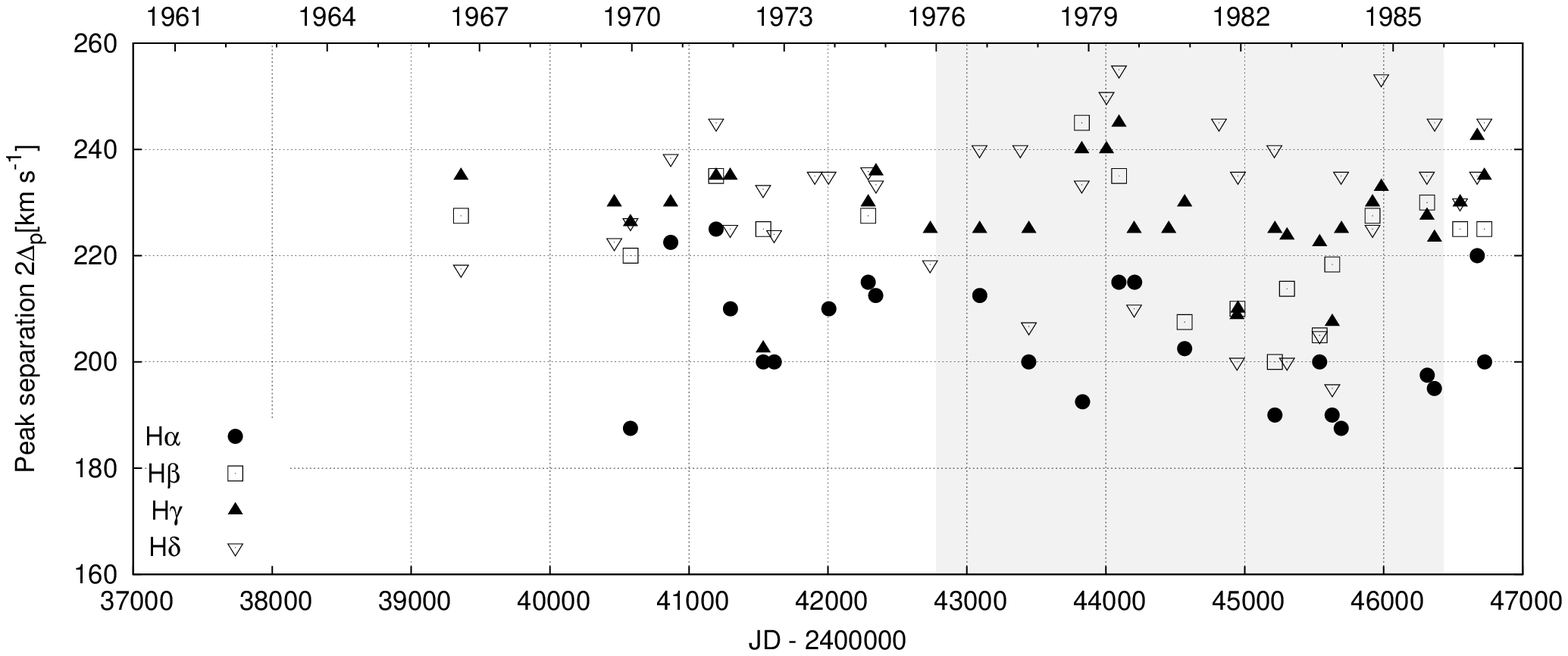}
  \end{center}
  \caption{Long-term variations of the full peak separation 2$\Delta_p$ [$\mathrm{km\;s^{-1}}$].}\label{fig:VarPS}
\end{figure*}

Full peak separations 2$\Delta _p$ are measured and the long-term variations are shown in figure \ref{fig:VarPS}.
It is seen that the $\Delta _p$ for every line indicates little correlation with the V/R variation.

The average values and the standard deviations are as follows:

$\Delta_p$ [$\mathrm{km\;s^{-1}}$]  = 101.5 $\pm$ 6.0 (H$\alpha$), 110.5 $\pm$ 6.1 (H$\beta$), 112.1 $\pm$ 15.3 (H$\gamma$) and 116.1 $\pm$ 11.3 (H$\delta$).

If we assume a circular disk around the equator, the peak separation gives the outer disk radius for a Keplerian disk, where each line is emitted.
Upon this assumption we get the outer radii of the disk as follows:

Outer radius/stellar radius = 7.5 (H$\alpha$), 6.3 (H$\beta$), 6.0 (H$\gamma$) and 5.9 (H$\delta$).

This infers that the size of emitting region is gradually getting smaller for higher members, even in case of elliptical disk.

\subsection{Central Core Components in the Emission Lines}\label{subsec:CentralCore}
Central core component is the part of line profile that is formed by the combination of emission line and shell absorption line. We have measured the line-profile parameters of this part in the H$\beta$, H$\gamma$ and H$\delta$ lines, such as radial velocity $V_c$ at the deepest point, line depth below the photospheric absorption  $D_c$ = CC' and full width of central absorption 2$\Delta_s$ = FG, as shown in figure \ref{fig:DefMeasure}. The results are shown in figure \ref{fig:VarCenCore}. We can see the observed features as follows:

\begin{figure*}
  \begin{center}
    \FigureFile(150mm,130mm){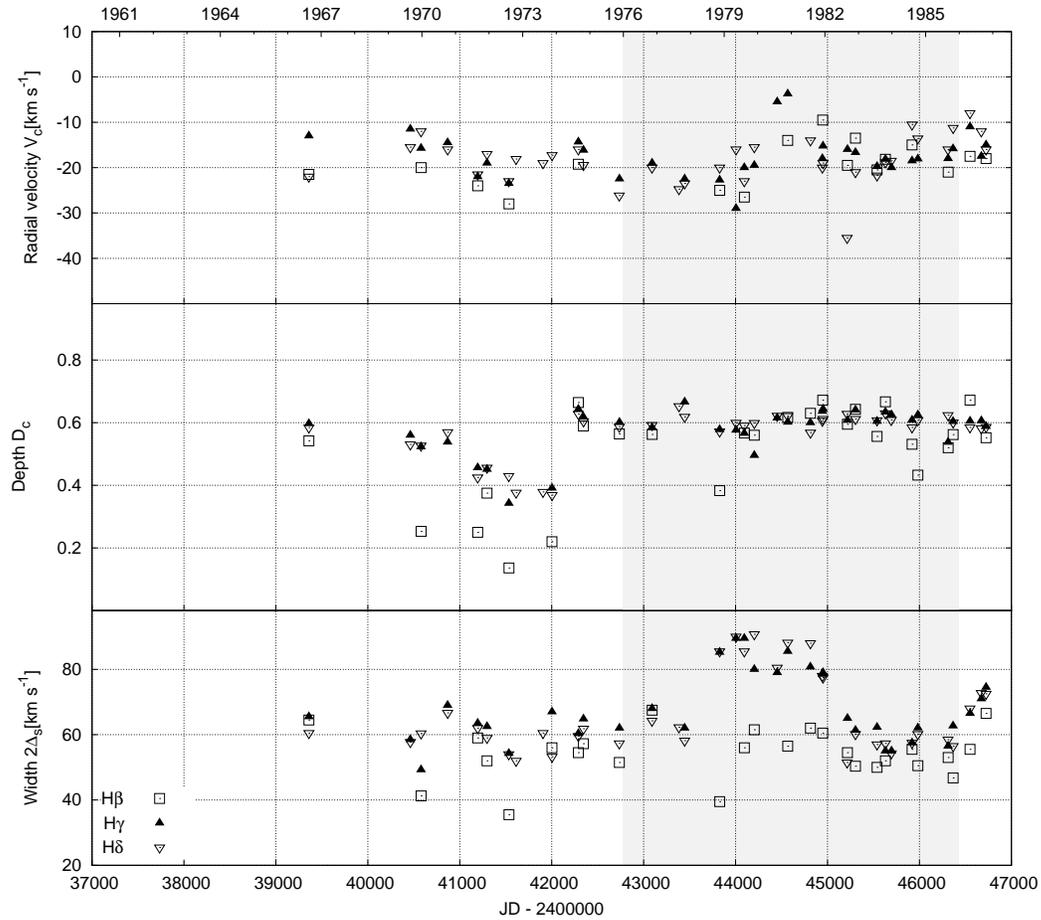}
  \end{center}
  \caption{Long-term variations of central core parameters, defined in figure \ref{fig:DefMeasure}. 
{\it Top panel} : Radial velocities.
{\it Middle panel} : Depths in unit of the residual intensity of the photospheric line center.
{\it Bottom panel} : Full width at half intensity.
}\label{fig:VarCenCore}
\end{figure*}

\begin{enumerate}
\item Radial velocities  $V_c$ exhibit different behaviors as compared to the V/R variations. Although data is lacking in H$\delta$ in 1980 -- 1981, there seems a lag in the time of reaching to the maximum among these lines with different amplitude. Similarly to the V/R variation, the H$\delta$ seems to have reached to the maximum first, and then the H$\gamma$ and the H$\beta$ lines followed. Time lag and velocity amplitude will be considered in section \ref{sec:RVofShell}.
\item Depths of the central core showed no particular variations during the V/R variations. We notice that there appeared a marked intensity decrease in 1972 -- 1973, prior to the V/R variation period. This feature will be considered in subsection \ref{subsec:Long_Shellar}.
\item Full widths of the central core absorption of the H$\gamma$ and H$\delta$ line exhibit a period of highly widened absorption as broad as 80 -- 90 $\mathrm{km\;s^{-1}}$ in 1978 -- 1981, as compared to that of the H$\beta$ line which is in 40 -- 60 $\mathrm{km\;s^{-1}}$. Since such large difference is difficult to explain by the difference of thermal motion, there should be some velocity structure inside the disk.
\end{enumerate}

\begin{figure*}
  \begin{center}
    \FigureFile(150mm,70mm){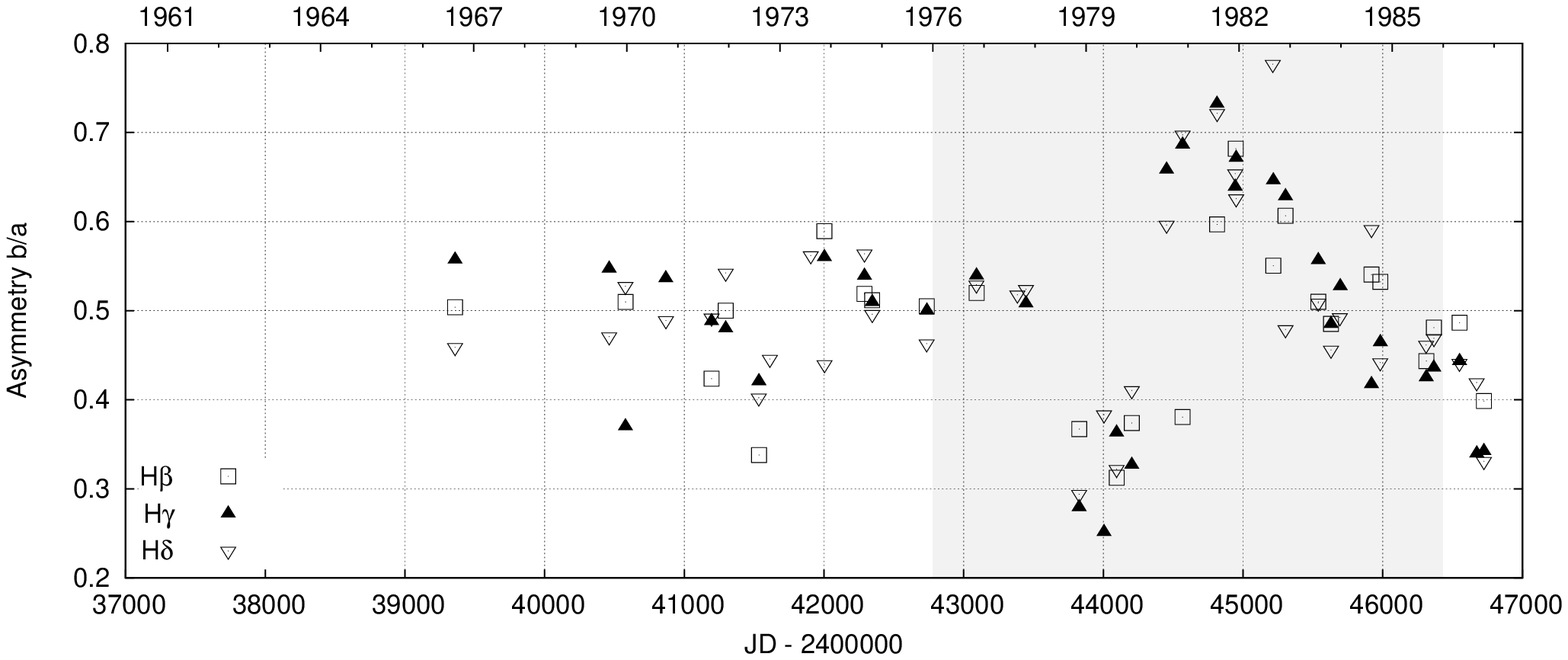}
  \end{center}
  \caption{Line asymmetry of the central core absorption in the H$\beta$ through the  H$\delta$ lines.
Asymmetry $b/a$ is defined in the text.
Large value above 0.5 is blue faded and below 0.5 is red faded.
}\label{fig:LineAsym}
\end{figure*}

In addition, we consider the line asymmetry of the central core absorption. We define the line asymmetry by $b/a$, where $a$ denotes the velocity difference between points F and G, and $b$ the difference between the velocities at point F and at CC'  (see figure \ref{fig:DefMeasure}). Then $b/a$ $<$ 0.5 indicates the red faded asymmetry and $b/a$ $>$ 0.5 the opposite. The long-term variations of the line asymmetry for the H$\beta$, H$\gamma$ and H$\delta$ are delineated in figure \ref{fig:LineAsym}. Clear correlation is seen with the V/R variations. The lines became red faded when V $>$ R (1980 to 1982) and then violet faded when V $<$ R (1978 to 1979) for all measured lines. Line asymmetry for the H$\beta$ line has already shown by \citet{HFCC87} in the same sense with ours. 

Line asymmetry is also shown by \citet{BACM87} in their trace of the V/R variation in the H$\beta$ line for HD184279 (V1294 Aql), where we can see the same sense of asymmetry as seen in figure \ref{fig:LineAsym}. In both stars, asymmetry is more prominent when V/R ratio is higher.

\subsection{Widths of Emission Line Wings}\label{subsec:WidWing}
The Balmer emission lines exhibit extended wings over the projected rotational velocity for the H$\alpha$ and higher members.
We have measured the widths of emission line wings from the line center both for the violet and red sides, $W_v$ and $W_r$ in $\mathrm{km\;s^{-1}}$ for the H$\beta$ line (D and E in figure \ref{fig:DefMeasure}).
The widths almost remain nearly constant before and during the V/R variation period and take the average velocities of

$W_v$ [$\mathrm{km\;s^{-1}}$] = 311.0 $\pm$ 20.5 and $W_r$ [$\mathrm{km\;s^{-1}}$] = 311.7 $\pm$ 31.0

These values are larger than the projected rotational velocity of 280 $\mathrm{km\;s^{-1}}$.
The situation is the same for the H$\gamma$ and the H$\delta$ lines (see figure \ref{fig:SamProf}).
These wide wings of emission profiles indicate that the circumstellar disk is not a detached ring, but a plain disk filled with gas.

\section{Recurrence of the V/R variation}\label{sec:VoverR2000s}
\subsection{Observations}\label{subsec:VoR2000Obs}
In order to see the state of V/R variations at present, an additional observation was made on 2010 October 22 at the Okayama Astrophysical Observatory.
By making use of the HIDES (High Dispersion Echelle Spectrograph), attached to the 188 cm reflector, we obtained a spectrum in a range of 4000 -- 7000 \AA.
Balmer lines are analyzed similarly with the method described in section \ref{sec:ObsData}.
The results of measurement are summarized in table \ref{tab:Emis2010}.
The definitions of physical quantities are the same as in figure \ref{fig:DefMeasure}.

\begin{table*}
  \caption{Emission line parameters on 2010 October 22}\label{tab:Emis2010}
  \begin{center}
    \begin{tabular}{ccccrccc}
      \hline
Line & 
\multicolumn{2}{c}{Intensity} & & & 
\multicolumn{3}{c}{Radial Velocity [$\mathrm{km\;s^{-1}}$]} \\

 & $I_v$ & $I_r$ & V/R & $EW_e$ [\AA] & $V_v$ & $V_r$ & $V_c$ \\
\hline \hline
H$\alpha$ & 2.38 & 3.40 & 0.70 & 21.37 & $-$123.5 & 104.0 & $-$20.0 \\
H$\beta$ & 0.54 & 0.91 & 0.59 & 2.98 & $-$144.0 & 104.0 & $-$18.0 \\
H$\gamma$ & 0.29 & 0.38 & 0.76 & 1.46 & $-$151.0 & 106.5 & $-$25.0 \\
H$\delta$ & 0.24 & 0.34 & 0.71 & 0.94 & $-$157.0 & 110.5 & $-$26.0 \\
     \hline
    \end{tabular}
  \end{center}
\end{table*}

\subsection{Long-Term V/R Variations}\label{subsec:Long_VoverR}
\begin{figure*}
  \begin{center}
    \FigureFile(150mm,70mm){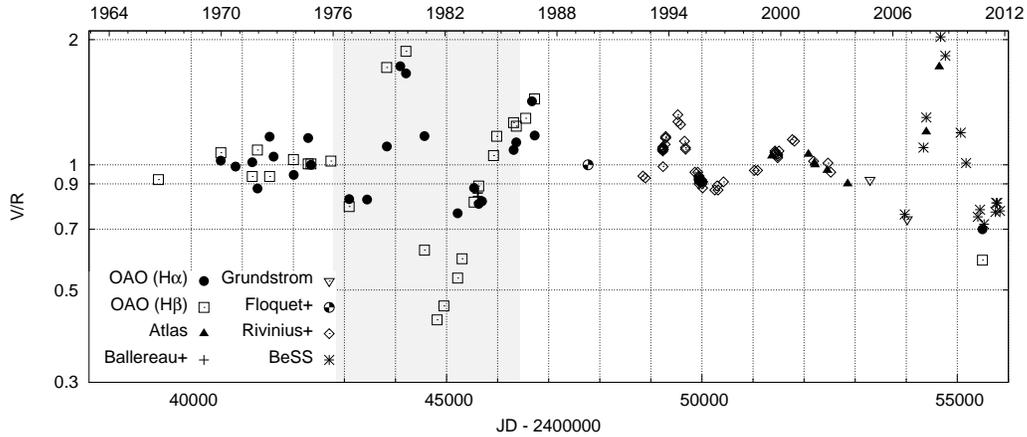}
  \end{center}
  \caption{Long-term V/R variations in the H$\alpha$ and H$\beta$ lines.}\label{fig:LongVar}
\end{figure*}

We have traced the long-term V/R variations of the H$\alpha$ and H$\beta$ lines up to 2011, using the data taken from \citet{BACM87}, \citet{G07}, \citet{FHHMZGHJKLSTN00}, \citet{RSB06} and the data sets of spectroscopic observations by Atlas\footnote{http://www.astrosurf.com/buil/us/bestar.htm} and BeSS database\footnote{http://basebe.obspm.fr/basebe/}. The survey result is shown in figure \ref{fig:LongVar}.

It is found that the V/R variation of 1976 -- 1986 did not ceased, but continued thereafter repeating irregular form of V/R variation. The V/R maximum appeared around in 1979, 1986, 1994, 2000 and 2008, so that the average period is approximately seven years. Among these repeated variations, the one that peaked in 2008 is most remarkable in its maximum and minimum V/R ratios. It seems a recurrence of the event of 1976 -- 1986 in its global behavior.

However, spectral features are different between the 1976 -- 1986 and the 2007 -- 2012 events. Our observation shows that EW Lac was in the state of V $<$ R in 2010 October. The emission equivalent width of the H$\alpha$ line (21 \AA) is much smaller than that in the previous event (around 50 \AA), and the peak separation (228 $\mathrm{km\;s^{-1}}$) is larger than that in the previous event (203 $\mathrm{km\;s^{-1}}$). The V/R ratios are almost the same among all of the emission lines. The implication of these differences will be discussed in section \ref{subsec:Time_Lag}.

\section{Shell Absorption Lines and Optical Depths}\label{sec:ShellVar}
\subsection{Method and Fittings}\label{subsec:Method}
\begin{figure}
  \begin{center}
    \FigureFile(80mm,50mm){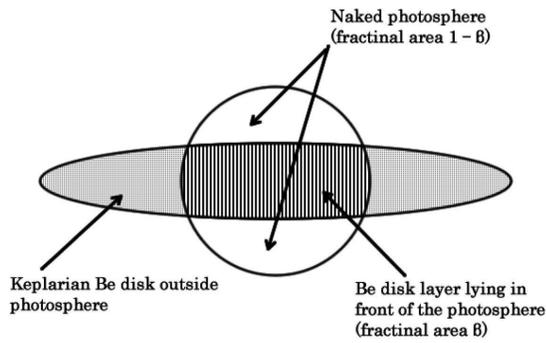}
  \end{center}
  \caption{
Equatorial view of the photosphere and circumstellar disk.
}
  \label{fig:Edge_on}
\end{figure}

Shell absorption lines are formed in the part of disk lying in front of the stellar photosphere.
This part is characterized by two parameters $\beta$ and $\tau$(H$\alpha$), where $\beta$ is the fractional area of this layer relative to the photospheric disk, and $\tau$(H$\alpha$) is the optical depth of the H$\alpha$ line in this layer.
In case of near edge-on star, this part of the disk is schematically illustrated in figure \ref{fig:Edge_on}.
The formula giving the central depths of the shell absorption lines in higher members of the Balmer series has been derived by \citet{KHA78} and applied to the analysis of Be-shell stars.
The same formula can also be derived as follows, based on a simple assumption that the shell lines are formed by pure absorption inside the disk.

Let $r_n$ and $r_*$ be the residual intensities at the centers of the line Hn and that of the photospheric absorption, relative to the nearby continuum $I_c$, respectively.
If we assume that there is no limb darkening on the photosphere, we have surface intensity $I_*(\lambda)$ for the part of naked photosphere, whereas we have $I_*(\lambda)e^{-\tau(\lambda)}$ for the part covered by the disk at any wavelength $\lambda$.
We then have $r_n=(I_*/I_c)\beta\exp [-\tau(\mathrm{Hn})]$ and $r_*=(I_*/I_c)(1-\beta)$, and the central depth $d_n$ of Hn, defined by $d_n=(r_* - r_n)/r_*$ is written as
\begin{equation}\label{eq:CentDepth}
d_n = \beta (1 - \exp [-\tau ( \mathrm{Hn} )]) = \beta (1 - \exp [-\omega _n \tau ( \mathrm{H\alpha} )]) ,
\end{equation}
where $\tau (\mathrm{Hn})$ denotes the optical depth for the line Hn, and
\begin{equation}\label{eq:Omega}
\omega _n = \frac{\nu _{2n}B_{2n}}{\nu _{23}B_{23}} = \frac{\tau ( \mathrm{Hn} )}{\tau ( \mathrm{H\alpha} )} .
\end{equation}
$B_{2n}$ is the Einstein coefficient for 2-n transition.
In deriving equation (\ref{eq:CentDepth}), we have assumed that there is no Balmer progression in the shell lines.
The security of this assumption will be discussed in section \ref{subsec:BalmerProg}.

\begin{figure*}
\centerline{
\begin{tabular}{cc}
\FigureFile(90mm,60mm){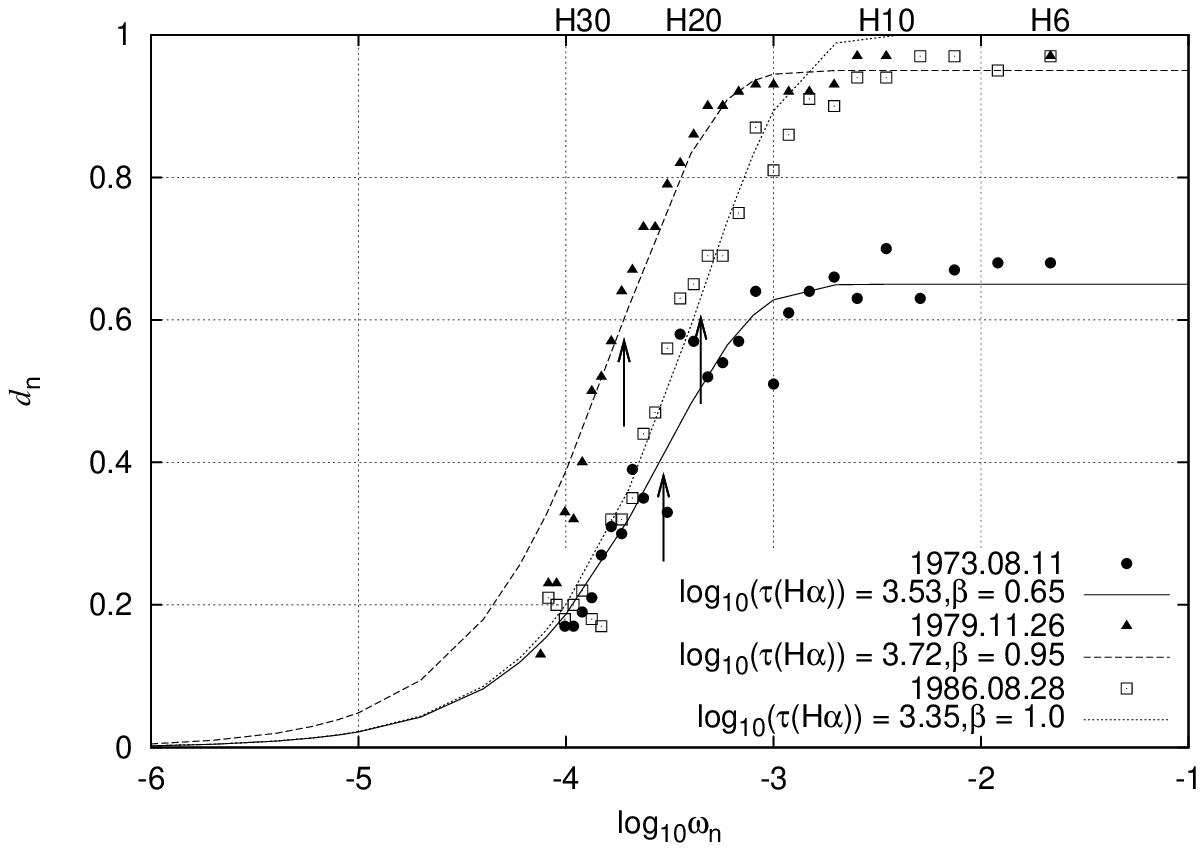} &
\FigureFile(90mm,60mm){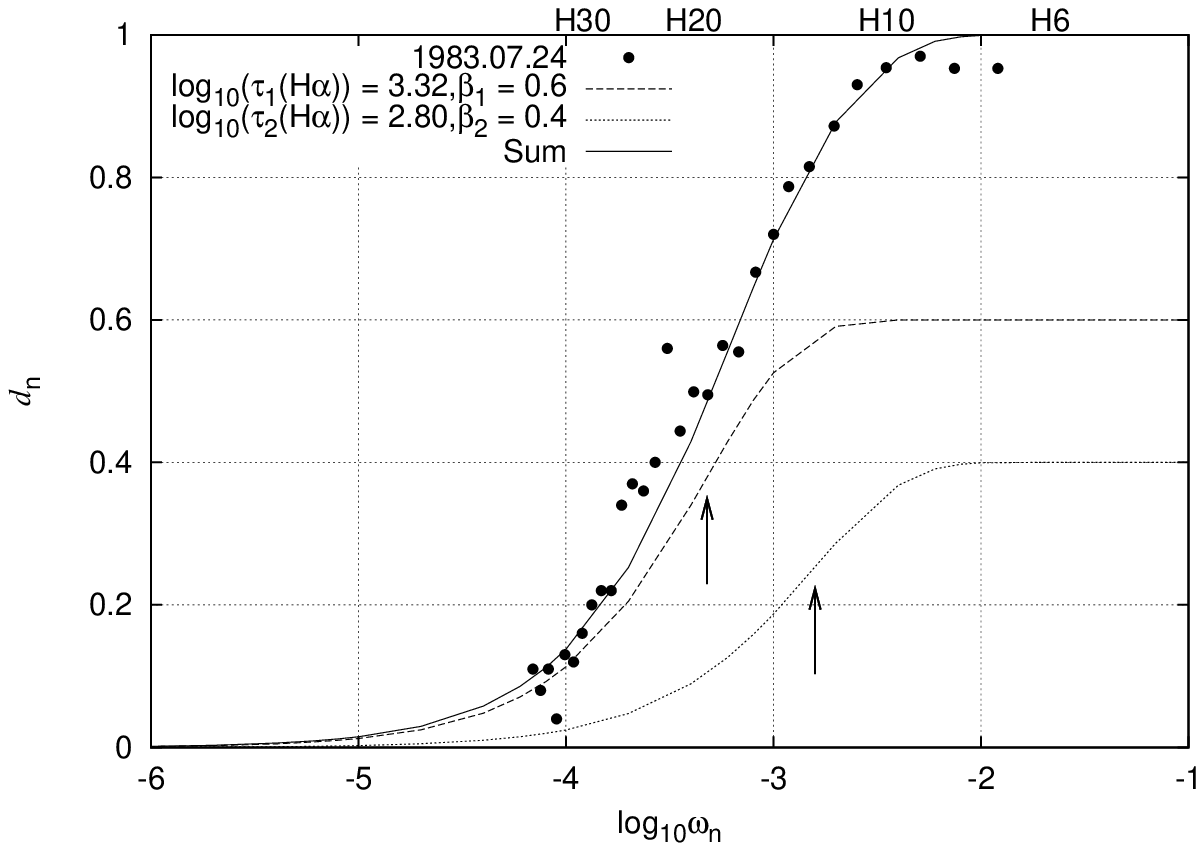} \\
\end{tabular}}
  \caption{
Fitting samples of central depth analysis for some epoch of EW Lac.
{\it Left panel} : Case of single-layer fitting.
{\it Right panel} : Case of double-layer fitting. 
In both panels, the date of observation and derived parameters are indicated.
The arrow in each theoretical curve indicates the position at which $\log \omega _n$ is zero on the equation (\ref{eq:CentDepth}).
}
  \label{fig:BT_fit}
\end{figure*}

By fitting the curves of equation (\ref{eq:CentDepth}) with observed points on the ($d_n$ -- $\log \omega _n$) diagram, we can get the values of $\tau$(H$\alpha$) and $\beta$ simultaneously by horizontal translation of the curve.
In some case, we need a combination of two curves of the formula (\ref{eq:CentDepth}) having two sets of $\tau _1$(H$\alpha$), $\beta _1$ and $\tau _2$(H$\alpha$), $\beta _2$, where $\tau _1$(H$\alpha$) $>$ $\tau _2$(H$\alpha$). The decomposition into two layers is made by a simple addition of the same formula (\ref{eq:CentDepth}). Some examples of single-layer and double-layer fitting are shown in figure \ref{fig:BT_fit}.

The values of $\tau$(H$\alpha$) and $\beta$ measured for EW Lac are partly given in table \ref{tab:Beta_Tau} (Full table is given in the electronic version). It contains the date, S/D (sign of adopted single- or double-layer fitting), and the values of $\tau$(H$\alpha$) and $\beta$ for the single and double layers, separately.

It is seen that the optical depth $\tau$(H$\alpha$) varies in a range from 2000 to 6000 for optically thick layer, and from 200 to 800 for optically thin layer.
In contrast, the parameter $\beta$ (single layer) or $\beta_1 + \beta_2$ (double layers) was always larger than 0.6, meaning that the disk is sufficiently extended in front of the photosphere.

\begin{table*}
\caption{The values of $\tau$(H$\alpha$) and $\beta$ for EW Lac.
Full table is given in the electronic version.}
\label{tab:Beta_Tau}
\begin{center}
\begin{tabular}{*{9}{c}}
\hline
Date	& JD2400000$+$	& S/D	& \multicolumn{2}{c}{Single layer}	& \multicolumn{4}{c}{Double layer} \\
	&	&	& $\tau$(H$\alpha$)	& $\beta$	& $\tau _1$(H$\alpha$)	& $\tau _2$(H$\alpha$)	&	$\beta=\beta _1 + \beta _2$	& $\beta _2$ \\ \hline \hline
1966/08/22	& 39359.210	& S	& 5210	& 0.90	\\
1969/08/29	& 40462.233	& S	& 2380	& 0.75	\\
1969/12/22	& 40577.893	& S	& 3050	& 0.85	\\
1969/12/23	& 40578.960	& S	& 3170	& 0.80	\\
1969/12/27	& 40582.902	& S	& 3270	& 0.85	\\
1970/10/08	& 40868.199	& S	& 4330	& 0.90	\\
1970/10/09	& 40869.184	& S	& 4920	& 0.85	\\
1971/09/01	& 41195.204	& S	& 4570	& 0.75	\\
1971/12/10	& 41296.028	& S	& 4570	& 0.78	\\
1972/08/04	& 41533.239	& S	& 1700	& 0.70	\\
1972/10/20	& 41611.068	& D	&	&	& 2690	& 380	& 0.70	& 0.20	\\
1972/10/22	& 41613.063	& D	&	&	& 3310	& 320	& 0.70	& 0.20	\\
1972/10/23	& 41614.092	& D	&	&	& 3540	& 360	& 0.70	& 0.20	\\
1972/10/24	& 41614.176	& D	&	&	& 3550	& 400	& 0.72	& 0.22	\\
1973/08/11	& 41905.167	& S	&	3390	& 0.65	&	&	&	\\ \hline
 \end{tabular}
\end{center}
\end{table*}

\subsection{Long-Term Variations}\label{subsec:Long_Shellar}
\begin{figure*}
  \begin{center}
    \FigureFile(150mm,90mm){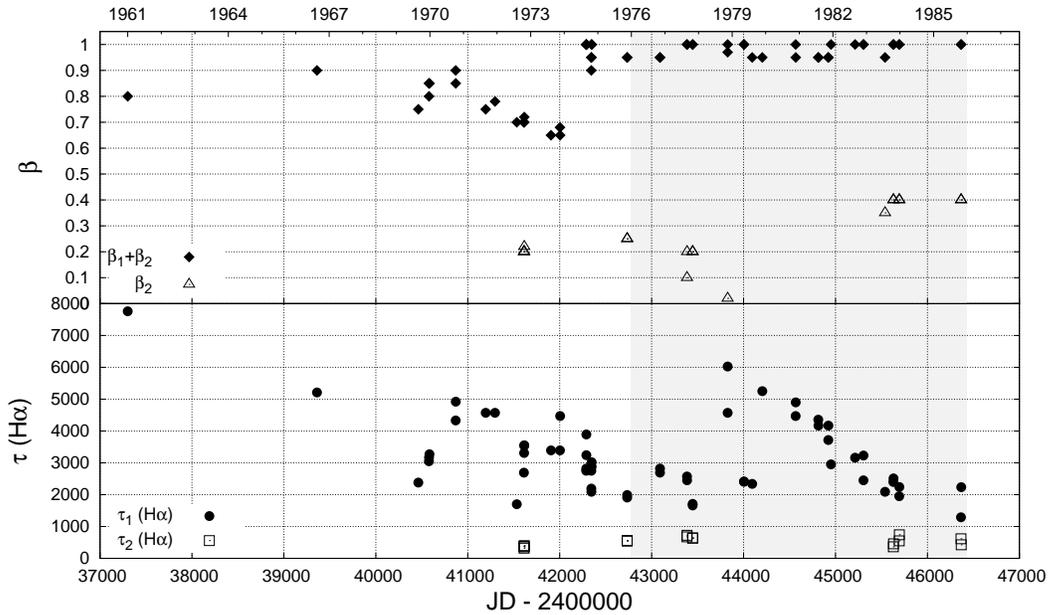}
  \end{center}
  \caption{
Long-term variations of $\beta$ (upper panel) and $\tau$(H$\alpha$) (lower panel).
The period of remarkable V/R variations is indicated by gray zone.
The values of 1961 are taken from \citet{O67}.
}
  \label{fig:Hist_BT}
\end{figure*}

The long-term variations of $\tau$(H$\alpha$) and $\beta$ are shown in figure \ref{fig:Hist_BT}.
These parameters evidently exhibit different behaviors as a whole as compared to the V/R variations or other emission line properties.

As seen in figure \ref{fig:Hist_BT}, $\tau$(H$\alpha$) showed a saw-teeth like variation: gradual decline, sudden increase and gradual decline in 1970's to 1980's.
It is to be noticed that the sudden increase in the middle of 1979 nearly corresponds to the epoch of the V/R maximum phase.
\citet{MZHGB98} also observed the sudden increase of the Balmer discontinuity in the same epoch after a long period of gradual decline.
This infers that the disk also becomes opaque in the Balmer continuum in this epoch.

It is also noticed in figure \ref{fig:Hist_BT} that, in some epochs, the disk is divided into optically thick and thin layers (double-layer fitting).
There seems to exist a tendency that the double layers appear when $\tau$(H$\alpha$) is relatively small, smaller than around 3500.
If we assume that EW Lac is seen nearly equator-on, we can suppose that the value of $\beta$ is a measure of the vertical height of the disk.
The disk is sufficiently vertically extended when $\tau$(H$\alpha$) is large, whereas, when $\tau$(H$\alpha$) is small, the disk vertically shrinks and the optically thick layer seems sandwiched between two optically thinner layers.

The value of $\beta$ has remained almost constant with its high level of 0.9 -- 1.0 throughout the period of the V/R variation.
One may notice that $\beta$ once dropped to around 0.65 in 1972 -- 73, and recovered to 0.9-level in 1974.
It is interesting to see that the central depths of emission profiles of the H$\alpha$ through the H$\delta$ lines, have markedly decreased in this period (subsection \ref{subsec:CentralCore}, figure \ref{fig:VarCenCore}).
The decrease of $\beta$ indicates the increase of the effective area of naked photosphere, so that this gives the explanation why the central depths of emission lines decreased.
It is because additional emission from the photosphere makes shallow the central depths apparently.

\subsection{Shell Line Depths along Balmer Series}\label{subsec:Bal_series}
Concerning the central depth analysis, \citet{K69} plotted the inverse central residual intensities of Balmer shell lines (line depth) against the series number for some epochs in 1965 September through December.
His purpose was the determination of electron density of the disk by making use of Inglis-Teller formula.
Kupo determined the value of the maximum number $n_m$ by extrapolating the observed line depths to the zero level of line depth.
He noticed that the observed residual intensities are often approximated by two broken lines, for which he interpreted by the existence of two layers with different electron density.

According to our present method, disappearance of shell absorption lines at highest members of the Balmer series is caused by the decrease of optical depths of the lines, which occurs before the lines are mutually blended by Stark effect.
This is apparent when we examine the line profiles of shell lines which are well separated in the highest members around H35 in case of EW Lac.
In weaker shell stars, shell lines disappear in series number much less than 30.

\begin{figure}
  \begin{center}
    \FigureFile(80mm,60mm){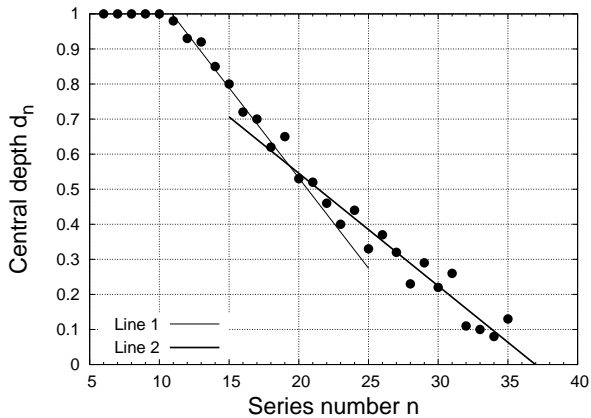}
  \end{center}
  \caption{
Shell line depths along the Balmer series number.
An example of double layer approximation at the date of 1975 August 30, is shown.
Optically thick layer is expressed by the regression line 2 and optically thinner layer is by the regression line 1.
The parameters of these layers are as follows: \\
\hspace*{3mm} $\beta _1$ = 0.80,  $\tau _1$ (H$\alpha$) = 2450  (for regression line 2) \\
\hspace*{3mm} $\beta _2$ = 0.20,  $\tau _2$ (H$\alpha$) = 720  (for regression line 1)
}
  \label{fig:Depth_Shell}
\end{figure}

From the view point of our central depth analysis, Kupo's one straight line indicates the case of one-layer disk with one-value of $\beta$, whereas the broken two lines correspond to the case of two-layer disk with two values of $\beta _1$ and $\beta _2$.
This is easily shown by producing the series number dependence of the central depths based on our measurements.
An example of broken lines is shown in figure \ref{fig:Depth_Shell}, in which our values of $\beta$ and $\tau$(H$\alpha$) for both broken lines are also indicated.
In this way we can explain Kupo's diagram in terms of the existence of single or double layers in the disk.

In Kupo's observations in 1965, one straight line appeared only in October, whereas two-broken lines are widely observed from September through December.
As stated in section \ref{subsec:Method}, two-layer state tends to appear when $\tau$(H$\alpha$) is smaller than around 3500.
Although we have no data of $\beta$ and $\tau$(H$\alpha$) in this period, we can guess that the disk in this year was optically thin, $\tau$(H$\alpha$) $\leq $ 3500, but variable in the value of $\beta$.

\subsection{Optical Depths of the Disk}\label{subsec:Shell_Depth}
Once $\tau$(H$\alpha$) is determined, we can deduce the optical depth for any Balmer lines $\tau$(Hn) through the value of $\omega _{2n}$.
The optical depths of the disk for the H$\beta$ through the H$\delta$ lines, relative to H$\alpha$, are shown in table \ref{tab:Opt_dep} for some cases of $\tau$(H$\alpha$).

\begin{table*}
\caption{Optical depth of the disk for the Balmer emission lines.}
\label{tab:Opt_dep}
\begin{center}
\begin{tabular}{*{9}{c}}
\hline
Line	& $\tau$(Hn)/$\tau$(H$\alpha$)$=\omega _{2n}$	& \multicolumn{7}{c}{Optical depths for some cases}	\\ \hline \hline
H$\alpha$	& 1.0 & 2000 & & & 4000 & & & 6000 \\
H$\beta$ & 0.16 & 320 & & & 640 & & & 960 \\
H$\gamma$ & 0.050 & 100 & & & 200 & & & 300 \\
H$\delta$ & 0.020 & 40 & & & 80 & & & 120 \\ \hline
 \end{tabular}
\end{center}
\end{table*}

\begin{figure*}
  \begin{center}
    \FigureFile(150mm,70mm){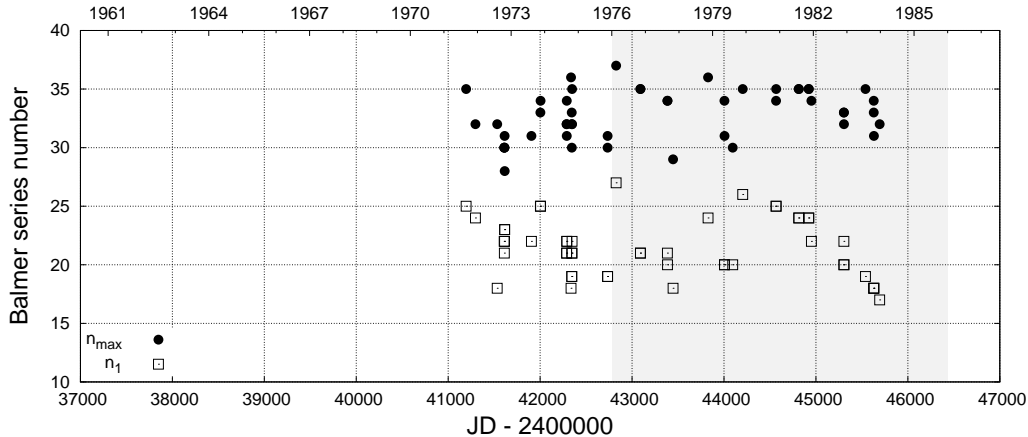}
  \end{center}
  \caption{
Maximum series number of recognizable Balmer lines $n_m$ and series number $n_1$  at which $\tau$(Hn)$=1$.
}
  \label{fig:Max_Ser}
\end{figure*}

In the same way we can obtain the optical depths of the shell lines through the values of $\omega _{2n}$.
In figure \ref{fig:Max_Ser} is shown the maximum series number $n_m$ of Balmer shell line recognized on the photographic plate, and the series number $n_1$ at which $\tau$(Hn)$=1$.
It is seen that both of $n_m$ and $n_1$ are almost unrelated with the V/R variations, and that the values of $n_1$ appear almost in between 16 and 25.
This infers that the average radial velocity $V (<$H16 -- H25$>)$, given below, is a good measure for indicating the gas motion inside the disk. 

\section{Radial Velocities of Shell Lines}\label{sec:RVofShell}
\subsection{Radial Velocities of the Line Centers}\label{subsec:RVofCent}
We measured the radial velocities of the central deep cores of line profiles, which correspond to the primary core noticed by \citet{HFCC87}.
Average heliocentric velocities of two cases of $V(<$H6 -- H15$>)$ and $V(<$H16 -- H25$>)$, along with the value of $V$(H20), are partly given in table \ref{tab:RV_Shell} (Full table is given in the electronic version).
The variation of radial velocity $V$(H20), measured by \citet{HFCC87} is in general coincidence with ours.

\begin{figure*}
  \begin{center}
    \FigureFile(150mm,60mm){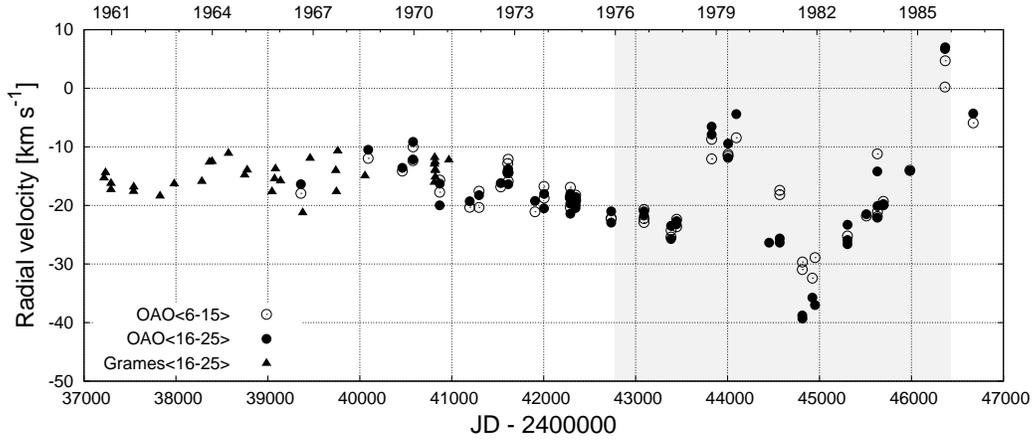}
  \end{center}
  \caption{
Long-term variation of the average radial velocities for higher members.
}
  \label{fig:RVofHigh}
\end{figure*}

The long-term variations of the average radial velocities $V(<$H6 -- H15$>)$ and $V(<$H16 -- H25$>)$ are shown in figure \ref{fig:RVofHigh}, along with the data of \citet{G72}.
Remarkable is the parallel relationship of radial velocities with the V/R variations given in figure \ref{fig:VoRPeakEW}.
This relationship infers that the V/R variations are closely connected with the internal gas motion of the disk.
It is notable that the radial velocity started to decrease in around 1973, before the commencement of remarkable V/R variation.
This infers that the disk matter in front of the star started to move outward in this early time, then gradually accelerated and turned to the inward motion in around 1978

Large amplitudes of the velocity curves make difficult to explain the V/R variation in terms of the contracting-expanding motion of gas elements in a circular disk.
This is because, if we calculate the distance of gas elements traveling during the expanding or contracting phase, the resultant distance is well over the whole size of the disk.
It is thus evident that the pulsation hypothesis (contraction/expansion) in a circular disk should be ruled out in case of EW Lac.
Close parallelism between the V/R and the radial velocity variations is suggestive of the existence of some non-circular disk, reflecting the projected velocity of some Keplerian motion, though the velocity structure of the disk is not clear.

It is also noticed that the radial velocities of the deepest points of central core for the H$\beta$ through H$\delta$ lines in around 1979, given in figure \ref{fig:VarCenCore}, disclose close relationship with $V(<$H6 -- H15$>)$ and $V(<$H16 -- H25$>)$ in the following two points:
First, the amplitude of variation is lower for lower member and higher for higher member, and secondly, the epoch of the maximum velocity is delayed in lower member.
These different behaviors may be attributed to the existence of emission components in the line profiles of the lower members and the difference of optical depths of the respective lines.
The latter implies that, since higher members are optically thin and formed in deeper layer, the disk should have larger gas motion in deeper layer as compared to its outer layer.

\begin{table*}
\caption{Radial velocities of Balmer shell lines. 
Full table is given in the electronic version.}
\label{tab:RV_Shell}
\begin{center}
\begin{tabular}{*{5}{c}}
\hline
Date	& JD 2400000$+$	&  $V(<$H6 -- H15$>)$	& $V(<$H16 -- H25$>)$	& $V$(H20)	\\
	& 	&  [$\mathrm{km\;s^{-1}}$]	& [$\mathrm{km\;s^{-1}}$]	& [$\mathrm{km\;s^{-1}}$]	\\ \hline \hline
1966/08/22	& 39359.210	& $-$17.9	& $-$16.4	& $-$15.0	\\
1968/08/23	& 40091.217	& $-$12.0	& $-$10.5	\\
1969/08/28	& 40462.233	& $-$14.1	& $-$13.6	& $-$15.0	\\
1969/12/22	& 40577.893	& $-$12.4	& $-$12.2	& $-$12.5	\\
1969/12/23	& 40578.960	& $-$10.0	& $-$9.2	& $-$10.0	\\
1970/10/09	& 40868.199	& $-$17.8	& $-$20.0	& $-$20.0	\\
1970/10/09	& 40869.184	& 	& 	& $-$15.0	\\
1970/10/09	& 40869.021	& $-$15.7	& $-$16.3	\\ \hline
 \end{tabular}
\end{center}
\end{table*}

\subsection{Balmer Progression}\label{subsec:BalmerProg}
\begin{figure*}
  \begin{center}
    \FigureFile(120mm,60mm){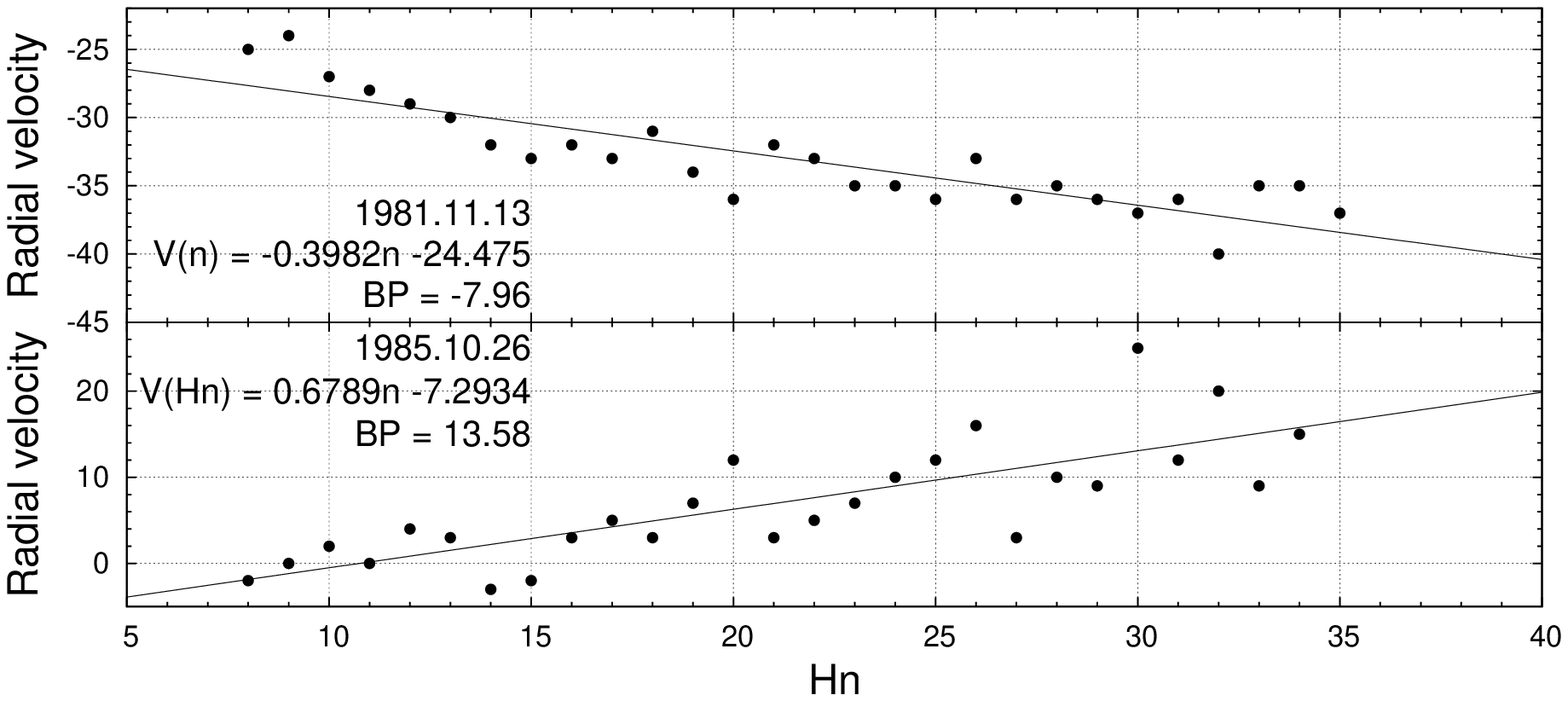}
  \end{center}
  \caption{
Examples of Balmer progression.
The date of observation, the regression line and derived value of $BP$ are indicated.
}
  \label{fig:SampleBP}
\end{figure*}

We have measured the amount of the Balmer progression, by creating a linear regression line on the radial velocity distribution along the Balmer series.
We simply define the Balmer progression ($BP$, in $\mathrm{km\;s^{-1}}$) by the formula
\begin{equation}\label{eq:balmer-prog}
BP =  V\mathrm{(H30)} - V\mathrm{(H10)} ,
\end{equation}
where $V$(Hn) denotes the radial velocities of shell lines Hn, read out on the linear regression line of the observed progression.
The values of  $V$(H10) and $BP$ are partly given in table \ref{tab:BPandRV10} (Full table is given in the electronic version).
Figure \ref{fig:SampleBP} illustrates a sample of Balmer progression measured in two epochs.

\begin{table*}
\caption{Balmer progression and radial velocity of the H10 line. 
Full table is given in the electronic version.}
\label{tab:BPandRV10}
\begin{center}
\begin{tabular}{*{4}{c}}
\hline
Date	& JD 2400000$+$	&  $BP$	& $V$(H10)	\\
	& 	& [$\mathrm{km\;s^{-1}}$]	& [$\mathrm{km\;s^{-1}}$]	\\ \hline \hline
1966/08/21	& 39359.210	& 1.1	& $-$17.0	\\
1966/08/22	& 40091.217	& $-$3.1	& $-$15.0	\\
1969/08/28	& 40462.233	& 1.5	& $-$12.0	\\
1969/12/23	& 40578.960	& 4.6	& $-$11.0	\\
1970/10/08	& 40868.199	& $-$1.7	& $-$18.0	\\
1970/10/09	& 40869.184 	& $-$0.8	& $-$17.0	\\
1971/09/01	& 41195.204	& 0.9	& $-$20.0	\\
1971/12/10	& 41296.028	& 1.9	& $-$21.0	\\ \hline
 \end{tabular}
\end{center}
\end{table*}

\begin{figure*}
  \begin{center}
    \FigureFile(150mm,70mm){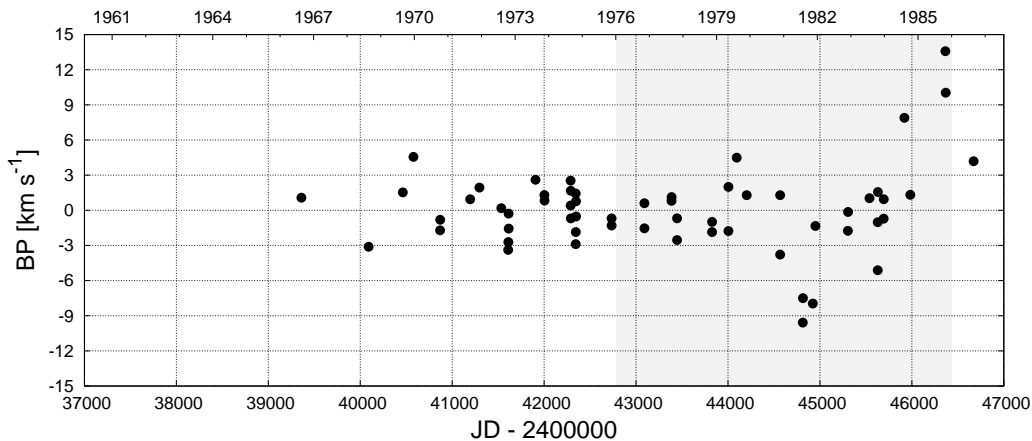}
  \end{center}
  \caption{
Long-term variation of the Balmer progression $BP$.
}
  \label{fig:Var_BP}
\end{figure*}

The long-term variation of $BP$ is shown in figure \ref{fig:Var_BP}.
In most of the observed period, $BP$ is confined within $\pm5$ $\mathrm{km\;s^{-1}}$.
Since thermal motion inside the disk is around 15 $\mathrm{km\;s^{-1}}$, provided the electron temperature is 10000 K or so.
The small value of $BP$ less than 15 $\mathrm{km\;s^{-1}}$ gives a security for the neglect of Balmer progression, adopted in section \ref{subsec:Method}.

Though amplitude is small, the $BP$ variation indicates some parallel correlation with the radial velocity variation of the shell lines given in figure \ref{fig:RVofHigh}.
This may reflects some effects of differential gas motion inside the disk.

\section{Physical Parameters of the Disk}\label{sec:Paras}
\subsection{Electron Density}\label{subsec:Elec_Dens}
By making use of the derived value of $\tau$(H$\alpha$), we can roughly estimate the distribution of electron density in the disk lying in front of the photosphere, provided the outer radius $R_d$ of the disk is known.
We shall tentatively adopt the outer radius $R_d = 7.5 R_*$ given in section \ref{subsec:PeakSep}, where $R_*$ denotes the radius of the photosphere.
Then we have (\cite{KL07}, definitions and references therein),
\begin{equation}\label{eq:def-tau}
d\tau ( \mathrm{H\alpha} ) =  K \frac{N_e^2(r)}{W(r)} dr ,
\end{equation}
and
\begin{equation}\label{eq:def-k}
K =  \frac{h\nu _{2n}}{4\pi} \frac{A_{c2} B_{2n}}{B_{2c}I_{2c}} ,
\end{equation}
where $K$ is a constant depending on the stellar temperature, and $K = 3.84 \times 10^{-33}$ for EW Lac ($T_{\mathrm{eff}}$ = 15800 K).
$W(r)$ is the dilution factor at the distance $r$ from the star center.
If we assume that the distribution of electron density $N_e(r)$ is given by a power law with index $\alpha$ as
\begin{equation}\label{eq:def-n}
N_e(r) =  N_{e0} \left( \frac{r}{R_*} \right) ^{-\alpha} ,
\end{equation}
where $N_{eo}$ denotes the electron density at the stellar surface.
Integrating equation (\ref{eq:def-tau}), we have
\begin{equation}\label{eq:def-tau2}
\tau ( \mathrm{H\alpha} ) =  K \int _{R_*}^{R_d} \frac{N_e^2(r)}{W(r)} dr .
\end{equation}
Since $\tau$(H$\alpha$) is already known, we can derive the distribution of electron density  $N_e(r)$, provided the values of $R_d$ and the index $\alpha$ are known.
Thus a sample of estimated electron density at the stellar surface is shown in table \ref{tab:Dens_elec}.
Notice that these values are not sensitive to the adopted outer radius $R_d$.

\begin{table*}
\caption{Estimated electron density at the stellar surface $N_{e0}$ [$\mathrm{cm^{-3}}$].}
\label{tab:Dens_elec}
\begin{center}
\begin{tabular}{*{4}{c}}
\hline
	& \multicolumn{3}{c}{$\tau$(H$\alpha$)} \\
$\alpha$	& 2000	& 4000	& 6000	\\ \hline \hline
2	& $6.91 \times 10^{11}$	& $9.78 \times 10^{11}$	& $1.20 \times 10^{12}$	\\
3	& $1.17 \times 10^{12}$	& $1.65 \times 10^{12}$	& $2.03 \times 10^{12}$	\\
4	& $1.56 \times 10^{12}$	& $2.20 \times 10^{12}$	& $2.70 \times 10^{12}$	\\ \hline
 \end{tabular}
\end{center}
\end{table*}

In table \ref{tab:Dens_elec}, it is noticed that the value of $N_{e0}$ at $\tau$(H$\alpha$) = 6000 is around 1.7 times larger than that at $\tau$(H$\alpha$) = 2000 for every value of $\alpha$.
It is also apparent that the density enhancement gives rise to little effects on the variation of equivalent widths of emission lines which remained almost unchanged in this period (figure \ref{fig:VoRPeakEW}).
This weak relationship may be caused by large optical depths of emission lines that are formed in the outer region of the equal line-of-sight velocity zone, and weakly affected by the density enhancement.
The relationship between density enhancement and formation of emission line profiles is the problem to be considered elsewhere, though the density enhancement is usually assumed as a cause of one-armed oscillations, or of some perturbations.

\subsection{Partial Mass of the Disk}\label{subsec:Mass_Disk}
By making use of the optical depth $\tau$(H$\alpha$) given in table \ref{tab:Dens_elec}, we can also estimate the partial mass of the disk $Mp$ that lies in front of the photosphere, by assuming this part as a cylinder with a base area of $\beta$-times the photospheric disk and a given optical depth $\tau$(H$\alpha$) for its height.
The result of estimation is given in table \ref{tab:Part_Mass}.
The increase of mass is around factor 1.7 from $\tau$(H$\alpha$) = 2000 to 6000. 
The partial mass slightly decreases for large value of the index $\alpha$.

As seen in figure \ref{fig:Hist_BT}, optical depth $\tau$(H$\alpha$) in front of the photosphere abruptly increased from around 2000 to 5000 during 400 days, i.e., from 1977 October through 1978 November.
This infers that the partial mass is also increased at the time.
A rough estimation, based on the values of table \ref{tab:Part_Mass} (Case of $\alpha$ = 3), gives the mass increasing rate of $3.3 \times 10^{-11} \; \mathrm{M_{\odot} \; yr^{-1}}$.
This value, however, does not necessarily indicate the mass loss rate from the photosphere.
If we assume an elliptical disk, massive part of the disk might have moved toward the front part of the photosphere.
 
\begin{table*}
\caption{Estimated mass of the disk lying in front of the photosphere.}
\label{tab:Part_Mass}
\begin{center}
\begin{tabular}{*{4}{c}}
\hline
	& \multicolumn{3}{c}{Partial mass $Mp$ in unit of solar mass} \\
$\alpha$	& 2000	& 4000	& 6000	\\ \hline \hline
2	& $6.65 \times 10^{-11}$	& $9.41 \times 10^{-11}$	& $1.15 \times 10^{-10}$	\\
3	& $6.40 \times 10^{-11}$	& $9.05 \times 10^{-11}$	& $1.11 \times 10^{-10}$	\\
4	& $5.77 \times 10^{-11}$	& $8.16 \times 10^{-11}$	& $9.99 \times 10^{-11}$	\\ \hline
 \end{tabular}
\end{center}
\end{table*}

\section{Photometric Variations and V/R Variations}\label{sec:Photo}
EW Lac is known as a remarkable photometric variable, and observations elucidate a large variety from short- to long-term variations.
In short-term variation within one day, evidence of multi-periodicity has widely observed \citep{P87,PRPJN93}, and some correlation with short-term V/R variation has been pointed out \citep{PS90}.
Long-term variations, much longer than one day, are mostly irregular, but reveal possible correlation with V/R variation in some case.

\begin{figure*}
  \begin{center}
    \FigureFile(150mm,140mm){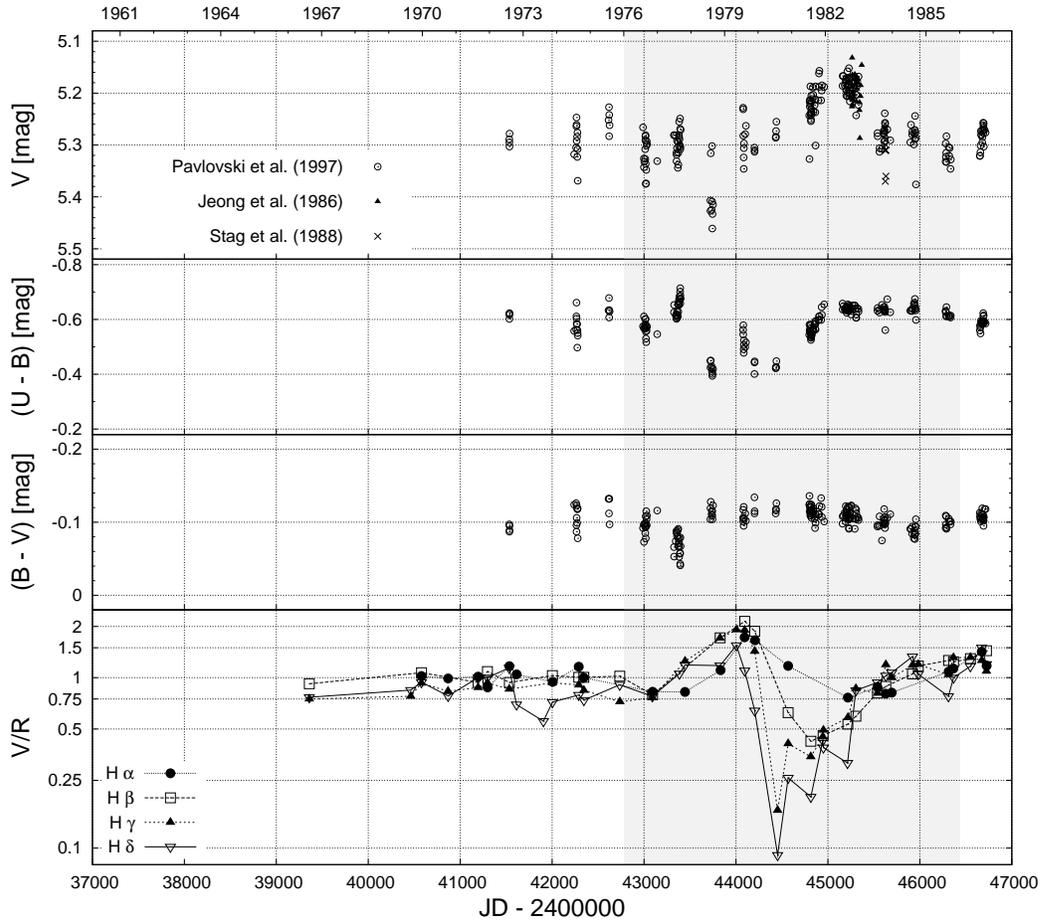}
  \end{center}
  \caption{
Light and color variations and V/R variations of EW Lac.
{\it Top panel} : $V$-magnitude, {\it Second panel} : $U- B$ color, {\it Third panel} : $B-V$ color, and {\it Bottom panel} : V/R variations.
}
  \label{fig:VoR_Photo}
\end{figure*}

In order to inspect the photometric behaviors of EW Lac during the V/R variation, we show in figure \ref{fig:VoR_Photo} the long term variations of $V$ magnitude and colors, based on the data of \citet{PHBKHKRS97}, \citet{JSN86} and \citet{SBFGGHHHIKKPPSSTZ88}.
For comparison, the V/R variations are also reproduced from figure \ref{fig:VoRPeakEW}.

It is seen that the V/R variation is related in some way with the light variation.
The relationship may be given as follows:

\begin{enumerate}
\item The remarkable brightening occurred in the latter half period of the V/R variation, while the darkening of light in around 1978 appeared just prior to the V/R maximum epoch.
In the epoch of the brightening in 1981 -- 1984, marked decrease of emission line intensities in the H$\alpha$ through the H$\delta$ lines were observed.
As seen in figure \ref{fig:VoRPeakEW}, equivalent width $EW_e$(H$\alpha$) decreased from 50 \AA \ level down to 40 \AA \ level in 1982 -- 1985.
\citet{BACM87} measured the equivalent widths of H$\alpha$ line in the epoch from 1983 September to October.
Although a marked short-term variation was seen in between 16.43 and 33.85 \AA \ , their average value is considerably lower than 50 \AA \ level.
On closer inspection of figure \ref{fig:RVofHigh}, we can see that the average radial velocity $V (<$H16-H25$>)$ reached a deep minimum of around $-35$ $\mathrm{km \;s^{-1}}$ in this epoch, indicating the existence of a strong outward gas motion in front of the photosphere.
In addition, the optical depth $\tau$(H$\alpha$) showed a small dip, $\Delta\tau$(H$\alpha$) $\sim$ 1000, in 1982 -- 1983 during its general decline phase as seen in figure \ref{fig:Hist_BT}.
These features suggest the stellar brightening has induced some spectroscopic changes in the disk.
One possible scenario may be that the stellar brightening has given rise to strong stellar wind, by which a part of disk mass has blown out, and as a result, emission line intensity has decreased markedly.
\item Prior to the V/R maximum phase, a remarkable darkening in $V$ band appeared accompanying some decrease of $U - B$ color in 1978.
It is also noticed that a marked dip around 0.05 magnitude occurred in 1977 (around JD 2443300).
The years 1977 -- 1978 correspond to the time when the gas motion in the inner part of the disk in front of the photosphere changed from outward to inward (figure  \ref{fig:RVofHigh}).
Temporal variations of brightness and colors in some irregular form seem to be connected with gas motions inside the disk.
\item Stellar brightness seems to have no direct effect to the optical structure of the disk, since the stellar brightening occurred while the optical depth $\tau$(H$\alpha$) is in its declining phase (figure \ref{fig:Hist_BT}).
\item \citet{PR91} claimed a close correlation between the brightness and radial velocity variations or shell activity for EW Lac.
In our observations, no such correlation was confirmed.
\end{enumerate}

\section{Discussions}\label{sec:Diss}
\subsection{Retrograde Nature of the V/R Variation}\label{subsec:Retro}
We suppose that the V/R variation is caused by the propagation of some density enhancement, not by the change of global shape (see sections \ref{subsec:PeakSep} and \ref{subsec:RVofCent}).
Then, by comparing the long-term variation of the V/R ratios in figure \ref{fig:VoRPeakEW}, and of the optical depth $\tau$(H$\alpha$) in figure \ref{fig:Hist_BT}, we can suppose that the V/R variation is of retrograde structure.

\begin{figure}
  \begin{center}
    \FigureFile(80mm,60mm){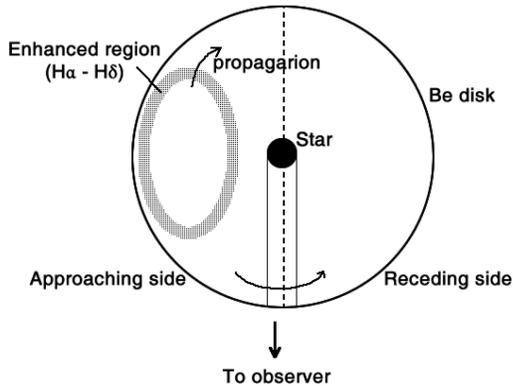}
  \end{center}
  \caption{
Schematic picture of the configurations of the density enhanced region at nearly the phase of V/R maximum in 1979.
The optical depth $\tau$(H$\alpha$) has already started its declining, so that enhanced region should propagate to the far side of the star.
}
  \label{fig:Conf_enhanced}
\end{figure}

First of all, notice that the optical depth reached its maximum in late 1978, and has started its nearly monotonous declining for the latter half of the V/R variation period (figure \ref{fig:Hist_BT}).
The V/R ratio showed its maximum in the middle of 1979.
When V/R ratio is in its maximum state, the enhanced region lies in the approaching side of the disk.
In case of prograde nature, the enhanced region should move to the front side of the disk, causing the increase of the optical depth.
In case of retrograde nature, this is in opposite, that is, the enhanced region moves to the behind side.
A schematic configuration of enhanced region for EW Lac is illustrated in figure \ref{fig:Conf_enhanced} at a phase of nearly V/R maximum.
After this phase, the optical depth $\tau$(H$\alpha$) gradually declined, and this implies that V/R variation is of retrograde structure in this star.

This retrograde structure of EW Lac seems particular among Be V/R variables.
\citet{THHS94} find the prograde nature for the case of $\beta^1$ Mon.
\citet{MSV97} carried out photometric observations for six stars (V923 Aql, V1294 Aql, $\gamma$ Cas, 48 Lib, MX Pup, $\zeta$ Tau) and found that these stars are compatible with prograde global disk oscillations based on the model of \citet{PSH92}.
\citet{O08} calculated 3D eccentric disk models of Be stars, and obtained the prograde modes for all reasonable disk conditions.
On the other hand, \citet{O91} mentioned in his one-armed oscillation model, that the fundamental modes and all of the overtones generally retrograde.
He also states that the prograde density structure found in $\beta^1$ Mon is difficult to explain by his model.
EW Lac may belong to this retrograde case.

\subsection{Time Lag and Spiral Structure}\label{subsec:Time_Lag}
In the V/R variations of EW Lac, two behaviors are to be noticed. One is the time lag for different Balmer lines, particularly in the latter half of the V/R variation period.
The second is the shortening of the duration of the V/R variation from H$\alpha$ toward  H$\delta$.
In table \ref{tab:DurVoR} we see that the total durations after the V/R maximum phase are 2340 (H$\alpha$), 1690 (H$\beta$), 1500 (H$\gamma$) and 1380 (H$\delta$) days, respectively.
The optical depth is smaller and the peak separation of Balmer emission is larger for higher members than for lower members.
These evidences indicate that higher members are formed in deeper layer than lower members.
This implies that the angular velocity of wave propagation is higher in deeper layer than in outer layer.
By combining time lag and phase shortening, we can suppose that the spiral structure possibly appeared in this period inside the disk.

\begin{figure}
  \begin{center}
    \FigureFile(72mm,52mm){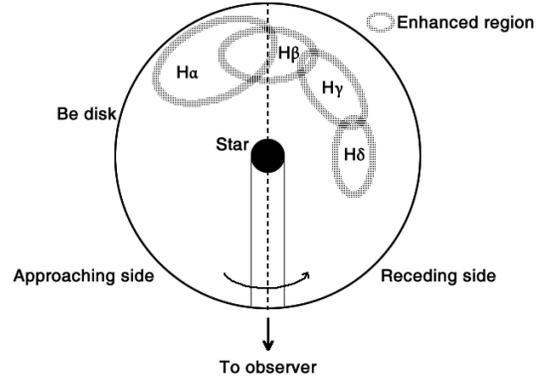}
  \end{center}
  \caption{
Configuration of enhanced regions at the phase of V/R minimum for H$\gamma$ and H$\delta$ lines, in the middle of 1980.
In this epoch, V $>$ R for H$\alpha$ and V $\sim$ R for H$\beta$.
}
  \label{fig:Conf_timelag}
\end{figure}

In figure \ref{fig:Conf_timelag} we depict a schematic configuration of the density enhanced regions of the lines H$\alpha$ through H$\delta$ at around the phase of V/R minimum for H$\gamma$ and H$\delta$ in the middle of 1980.
After the phase of V/R maximum shown in figure \ref{fig:Conf_enhanced}, the enhanced regions are separated in the sense that higher members propagate faster than lower members, thus the spiral structure is formed as shown in figure \ref{fig:Conf_timelag}.
We can trace this spiral structure up to around 1983.
In 1984, the enhanced regions successively dispersed before entering the front side of the photosphere, so that the V/R ratios has become unity and the optical depth decreased down to the value of $\tau$(H$\alpha$) $\sim$ 2000.

\citet{O91} mentioned, in his one-armed oscillation model, that, when one-armed waves are excited, the most perturbed region is likely to be in the innermost part of the disk, and propagate outward with a group velocity depending on their frequencies.
\citet{COBSRBBH09} calculated 2D global oscillation model in the disk of binary star, $\zeta$ Tau, and showed an appearance of a pair of or single spiral-like structure.
\citet{O91} also mentioned that the propagation of such perturbed region is merely a transient one.
Spiral structure in EW Lac appeared only in latter half of the V/R variation period, so that the spiral structure of EW Lac may be corresponding to the transient phenomena predicted by Okazaki.
\citet{RSB06} claim the finding of phase lag and helical structure of the density waves in the long-term V/R variables, maybe including 48 Lib, though the details are not shown.

In the V/R variation period started in 2007, it may be interesting to see whether similar spiral structure appeared or not after the V/R maximum phase.
Although observational data are quite limited (Section \ref{subsec:VoR2000Obs}), there is one clue to guess the absence of spiral structure.
That is, the V/R ratios have shown similar values for all of emission lines at the epoch of 2010 October, several hundred days after the V/R maximum, suggesting the lacking of time lag as compared with the first event.
Time lag seems to be closely related to the formation of spiral structure.

\subsection{Weak Correlation and Possible Binary Effect}\label{subsec:Binary_effect}
As shown in section \ref{sec:SpecVar}, the V/R variation indicates weak correlations with peak radial velocities, peak separations of emission lines, and emission equivalent widths.
These weak correlations suggest the structure of the disk truncated at some radius from the star center.
One possibility for such truncation is to assume the existence of unseen small-mass companion around the primary.
According to the stable orbit theory of \citet{HG70}, gas particles around the primary make some stable orbits, and the maximum stable orbit (MSPO) defines the outer boundary of circumstellar mass around the primary.
When the mass ratio is sufficiently small such as less than 5 percent, MSPO gives rise to two effects: one is the existence of outer boundary of the disk, and second is small eccentricity of the disk.
These effects might explain the existence of weak correlation among physical parameters when applied to EW Lac.
This possibility, however, remained as a future problem.

\section{Summary}\label{sec:Summ}
EW Lac entered the repeating V/R variation phase in 1976 with approximate period of seven years.
Among those, remarkable V/R variation has been observed in 1976 -- 1986, and in the time started in late 2000's.

We have mainly investigated the spectroscopic behaviors of EW Lac during the first V/R variation event.
We analyzed emission lines and shell absorption lines in the Balmer series.
In the analysis of emission lines, we found that the V/R variations of the H$\alpha$ through the H$\delta$ lines are characterized by time lag and shortening of phase duration from lower to higher Balmer members.
Time lag is conspicuous after the V/R maximum and the V/R ratio reached its minimum value in the order from H$\delta$ to H$\alpha$.
The phase duration is longest in H$\alpha$ and shortest in H$\delta$.
We also found weak correlations between V/R variation and physical parameters such as peak velocities, emission equivalent widths and peak separations.

We have then analyzed shell absorption lines for higher members of the Balmer series, which are formed in the disk lying in front of the photosphere.
Our finding is as follows:

\begin{enumerate}
\item Disk is sufficiently opaque for the emission lines and the optical depth is as high as 2000 to 6000 in the H$\alpha$ line.
Optical depth is getting lower and becomes unity in around H15 -- H25.
\item Long-term variation of the optical depth of H$\alpha$ displays quite different behavior as compared to the V/R variation.
However, the epoch of maximum depth in 1979 nearly coincides with the epoch of V/R maximum.
\item The effective area of the disk covering the photosphere is high during the V/R variation period, implying that the disk is vertically extended in front of the photosphere.
\item Average radial velocity of the shell absorption lines showed long-term variation parallel with that of the V/R variation.
This implies that the V/R variations are associated with the internal velocity structure of the disk.
\item Electron density and partial mass in front of the photosphere are estimated in some cases.
Electron density showed the change around 1.5 $\sim$ 1.7 times during the V/R variation.
\end{enumerate}

Based on the analysis of emission and shell absorption lines, we have considered the disk structure as follows:

\begin{enumerate}
\item The V/R variation is found to have retrograde structure.
\item Spiral structure is likely observed inside the disk in the latter half of the V/R variation period.
\item Possible effect of unseen small-mass companion is suggested to explain the weak correlation between V/R variation and other physical parameters.
Binary effect, however, remained as a future problem.
\item Close relationship between photometric variation and the V/R variation is pointed out.
The remarkable brightening of the star in the later stage of the V/R variation may have induced a mass loss from the disk, causing a marked decrease of emission line intensity.
\end{enumerate}

The V/R variation appeared in 2007 is just like a recurrence of the 1976 -- 1986 event.
Our observations in 2010 showed low emission intensities, and similar values of V/R ratios for H$\alpha$ through H$\delta$, several hundred days after the V/R maximum epoch.
This suggests the absence of time lag.

EW Lac may be the first star that time lag and variation of phase duration in the V/R variation have been observed.
Present study yields important constraints for the modeling of the V/R variation of this star.

\bigskip

The authors are grateful to the staff of the Okayama Astrophysical Observatory for the use of archive data of EW Lac and their kind hospitality. In our analysis we used a part of spectral plates obtained by R. Hirata and T. Horaguchi to whom we express our thanks.
We express our thanks to Dr. P. Harmanec, Astronomical Institute, Praha, Czech Republic, for sending the detailed photometric data of this star.
We express our gratitude for kind comment of anonymous referee.


\begin{thebibliography}{}

\bibitem[Ballereau et al.(1987)]{BACM87}
Ballereau, D., Alvarez, M., Chauville, J., \& Michel, R. 1987, Rev. Mexicana Astron. Astrof., 15, 29
\bibitem[Carciofi et. al.(2009)]{COBSRBBH09}
Carciofi, A. C., Okazaki, A. T., Le Bouquin, J. -B., Štefl, S., Rivinius, Th., Baade, D., Bjorkman, J. E., Hummel, C. A. 2009, \aap, 504, 915
\bibitem[Chauville et al.(2001)]{CZBMCG01}
Chauville, J., Zorec, J., Ballereau, D., Morrell, N., Cidale, L. \& Garcia, A. 2001, \aap, 378, 861
\bibitem[Floquet et al.(2000)]{FHHMZGHJKLSTN00}
Floquet, M. et al. 2000, \aap, 362, 1020
\bibitem[Granes(1972)]{G72}
Granes, P. 1972, \aap, 19, 224
\bibitem[Grundstrom(2007)]{G07}
Grundstrom, E. D. 2007, Ph.D. thesis, Georgia State University
\bibitem[H{\'e}non \& Guyot(1970)]{HG70}
H{\'e}non, M. \& Guyot, M 1970, Periodic Orbits Stability and Resonances, Proceedings of a Symposium, held at the University of Sao Paulo, Brasil, September 4-12, 1969. Edited by G.E.O. Giacaglia. Dordrecht: D. Reidel Publishing Company, 1970, p.349
\bibitem[Hubert et al.(1987)]{HFCC87}
Hubert, A. M., Floquet, M. Chauville, J. \& Chambon, M. T. 1987, \aaps, 70, 443
\bibitem[Jeong et al.(1986)]{JSN86}
Jeong, J. H., Suh, C. W. \& Nha, I.-S. 1986, \apss, 119, 73
\bibitem[Kogure(1975)]{K75}
Kogure, T. 1975, \pasj, 27, 165
\bibitem[Kogure et al.(1978)]{KHA78}
Kogure, T., Hirata, R. \& Asada, Y. 1978, \pasj, 30, 385 
\bibitem[Kogure and Leung(2007)]{KL07}
Kogure, T. \& Leung, K. C. 2007, Astrophysics of Emission Line Stars (Springer), Ch. 4 \& 5
\bibitem[Kogure and Suzuki(1984)]{KS84}
Kogure, T. \& Suzuki, M. 1984, \pasj, 36, 191
\bibitem[Kupo(1969)]{K69}
Kupo, I. D. 1969, Trudy Astrofiz. Inst. (Alma-Ata), 14, 3 
\bibitem[Kurucz(1979)]{K79}
Kurucz, R. L. 1979, \apjs, 40, 1
\bibitem[Mennickent et al.(1997)]{MSV97}
Mennickent, R. E., Sterken, C. \& Vogt, N. 1997, \aap, 326, 1167
\bibitem[Moujtahid et al.(1998)]{MZHGB98}
Moujtahid, A., Zorec, J., Hubert, A. M., Garcia, A. \& Burki, G. 1998, \aaps, 129, 289
\bibitem[Ogilvie (2008)]{O08}
Ogilvie, G. I. 2008, \mnras, 388, 1372.
\bibitem[Okazaki(1991)]{O91}
Okazaki, A. T. 1991, \pasj, 43, 75
\bibitem[\"{O}zemre(1967)]{O67}
\"{O}zemre, K. 1967, Annales d'Astrophysique, 30, 495
\bibitem[Papaloizou et al.(1992)]{PSH92}
Papaloizou, J. C., Savonije, G. J. \& Henrichs, H. F. 1992, \aap, 265, L45
\bibitem[Pavlovski(1987)]{P87}
Pavlovski, K. 1987, \apss, 134, 317
\bibitem[Pavlovski and Schneider(1990)]{PS90}
Pavlovski, K. \& Schneider, H. 1990, \aap, 228, 361
\bibitem[Pavlovski \& Ru\v{z}i\'{c}(1991)]{PR91}
Pavlovski, K. \& Ru\v{z}i\'{c}, Z. 1991, in Proc. ESO Workshop, Rapid variability of OB stars: Nature and diagnostic. p.191
\bibitem[Pavlovski et al.(1993)]{PRPJN93}
Pavlovski, K., Ru\v{z}i\'{c}, \v{Z}., Pavlovic, M., Jeong, J.~H. \& Nha, I.-S. 1993 \apss, 200, 201
\bibitem[Pavlovski et al.(1997)]{PHBKHKRS97}
Pavlovski, K., Harmanec, P., Bozic, H., Koubsky, P., Hadrava, P., Kriiz, S., Ru\v{z}i\'{c}, Z. \& Stefl, S. 1997, \aaps, 125, 75
\bibitem[Poeckert(1980)]{P80}
Poeckert, R. 1980, Publ. Dom. Ap. Obs., 15, 327
\bibitem[Porter and Rivinius(2003)]{PR03}
Porter, J. M. \& Rivinius, T. 2003, \pasp, 115, 1153
\bibitem[Rivinius et al.(2006)]{RSB06}
Rivinius, Th., Stefl, S \& Baade, D. 2006, \aap, 459, 137
\bibitem[Stagg et al.(1988)]{SBFGGHHHIKKPPSSTZ88}
Stagg, C. R. et al. 1988, \mnras, 234, 1021
\bibitem[Suzuki and Kogure(1986)]{SK86}
Suzuki, M. \& Kogure, T. 1986, \apss, 119, 69
\bibitem[Telting et al.(1994)]{THHS94}
Telting, J. H., Heemskerk, M. H. M., Henrichs, H. F. \& Savonije, G. J. 1994, \aap, 288, 558

\end{thebibliography}
\end{document}